\documentclass[letterpaper]{nature}

\usepackage{amsmath}
\usepackage{amsfonts}
\usepackage{graphicx}
\usepackage{amssymb}
\usepackage{caption}
\usepackage{afterpage}
\usepackage{subcaption}
\usepackage{url}
\usepackage{enumitem}
\usepackage{fullpage}
\usepackage{cite}
\usepackage{citesupernumber}
\usepackage{lineno}
\usepackage{anyfontsize}
\usepackage{booktabs}
\usepackage{dcolumn}
\newcolumntype{d}[1]{D..{#1}} 
\usepackage{pbox}
\usepackage{multibib}
\newcites{supp}{$\qquad$}
\usepackage{color}
\usepackage{xcolor}

\usepackage{url}
\usepackage{breakurl}
\usepackage[breaklinks]{hyperref}

\newcommand{\hide}[1]{}

\newcommand{\xhdr}[1]{\vspace{1.7mm}\noindent{{\bf #1.}}}
\newcommand{\disc}[1]{#1}

\newcommand{\numdaysThirty}{248,266 }
\newcommand{\numdaysNinety}{595,803 }
\newcommand{\numcities}{1,609 }
\newcommand{\numusers}{5,424}
\newcommand{\nummoves}{7,447 }
\newcommand{\internummoves}{31,034 }

\newcommand{\numtotalusers}{2,112,288}
\newcommand{\pctchangeCHIPHI}{11.2\% }
\newcommand{\millionmoreCHIPHI}{36 }
\newcommand{\pctchangeNYC}{14.5\% }
\newcommand{\millionmoreNYC}{47 }
\newcommand{\pctnowmeetguidelines}{18\% }
\newcommand{\millionnowmeetguidelines}{58 }
\newcommand{\pctMVPAafter}{42.5\%}
\newcommand{\pctMVPAbefore}{21.5\%}

\newcommand{\participants}{subjects}
\newcommand{\participant}{subject}
\newcommand{\Participant}{Subject}
\newcommand{\Participants}{Subjects}
\newcommand{\ParticipantsApp}{Subjects'}

\newcommand{\eg}{\emph{e.g.}}
\newcommand{\ie}{\emph{i.e.}}

\newif\ifabstractheaders
\abstractheadersfalse

\newif\ifheaders
\headersfalse

\newif\iffigures
\figurestrue

\makeatletter
\let\@fnsymbol\@arabic
\makeatother
\makeatletter
\@fpsep\textheight
\makeatother

\author
{Tim Althoff$^{1}$, Boris Ivanovic$^{2}$, Jennifer L. Hicks$^{3}$, Scott L. Delp$^{3,4}$, Abby C. King$^{5,6}$, Jure Leskovec$^{7,8}$\\ %
\vspace{-1.5\baselineskip}\\\normalsize{$^{1}$Allen School of Computer Science \& Engineering, University of Washington} \\
\normalsize{$^{2}$NVIDIA Research, Santa Clara, CA} \\
\normalsize{$^{3}$Department of Bioengineering, Stanford University} \\
\normalsize{$^{4}$Department of Mechanical Engineering, Stanford University} \\
\normalsize{$^{5}$Department of Epidemiology \& Population Health, Stanford University School of Medicine}\\
\normalsize{$^{6}$Stanford Prevention Research Center, Department of Medicine, Stanford University School of Medicine}\\
\normalsize{$^{7}$Department of Computer Science, Stanford University} \\
\normalsize{$^{8}$Chan Zuckerberg Biohub, San Francisco, CA}\\
\vspace{-1.5\baselineskip}\\
}
\title{\large
Countrywide natural experiment reveals impact of built environment on physical activity
}
\begin{document}

\spacing{1.4} %

\maketitle

{%\linenumbers
\begin{abstract}

\clearpage 
\section*{\Large Abstract}

While physical activity is critical to human health,
most people do not meet recommended guidelines\cite{2018physicalactivityguidelines,Bull2020WHO2020Guidelines}.
More walkable built environments have the potential to increase activity across the population\cite{althoff2017large,kohl2012pandemic,Sallis2016,giles2016city,deweerdt2016urban,graham2013designs}.
However, previous studies on the built environment and physical activity have led to mixed findings, 
possibly due to methodological limitations such as small cohorts, few or single locations, over-reliance on self-reported measures, and cross-sectional designs\cite{deweerdt2016urban,goenka2016our,Sallis2016,Ding2018,prince2022}. 
Here, we address these limitations by leveraging a large U.S. cohort of smartphone users (N=\numtotalusers) to evaluate within-person longitudinal behavior changes that occurred over \numdaysThirty days of objectively-measured physical activity across 7,447 relocations among 1,609 U.S. cities. 
By analyzing the results of this natural experiment, which exposed individuals to differing built environments, we find that increases in walkability are associated with significant increases in physical activity after relocation (and vice versa).
These changes
hold across subpopulations of different genders, age, and body-mass index (BMI), and are sustained over three months after moving. %
The added activity observed after moving to a more walkable location is predominantly composed of moderate-to-vigorous physical activity (MVPA), which is linked to an array of associated health benefits across the life course\cite{2018physicalactivityguidelines}. 
A simulation experiment demonstrates that substantial walkability improvements (i.e., bringing all US locations to the walkability level of Chicago or Philadelphia) may lead to 10.3\% or 33 million more Americans meeting aerobic physical activity guidelines. 
Evidence against residential self-selection confounding is reported.
Our findings provide robust evidence supporting the importance of the built environment in directly improving health-enhancing physical activity, in addition to offering potential guidance for public policy activities in this area.

\end{abstract}
\clearpage

\begin{figure}[tb]
\vspace{-5mm}
\centering
\includegraphics[width=\textwidth]{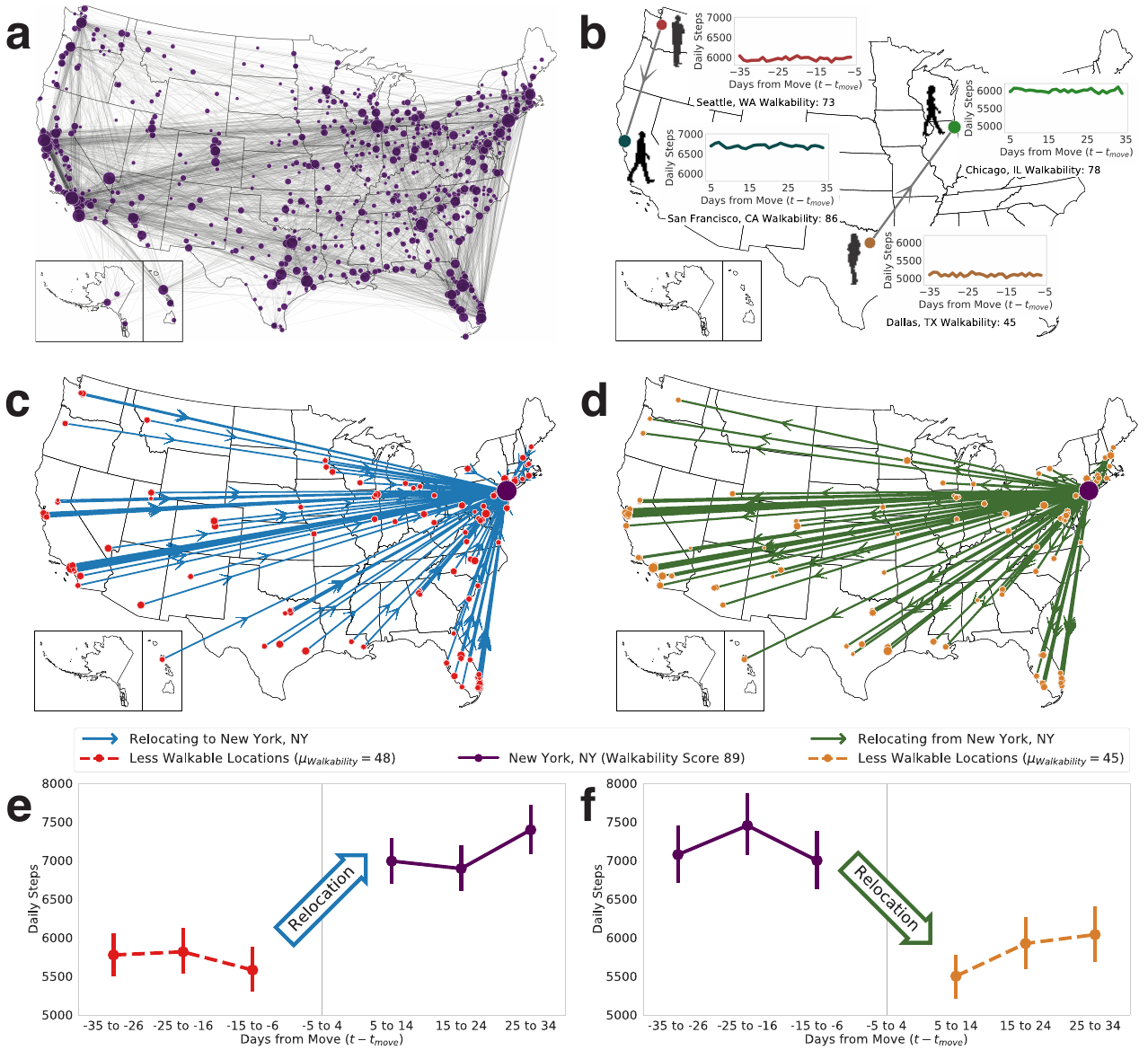} 
\vspace{-10mm}
\caption{
\textbf{Physical activity levels undergo significant changes following relocation between U.S. cities of different walkability levels.}
\textbf{a, }
During the observation period, \numusers~\participants~relocated \nummoves times between \numcities U.S. cities. 
Circle area is proportional to the square root of the number of relocations to and from the city. 
\textbf{b, }
\Participants' physical activity levels were tracked through smartphone accelerometry over several months before and after relocation, creating a countrywide study of \nummoves quasi-experiments.
\textbf{c,e }
Physical activity of \participants~moving from less walkable locations to New York City, \textbf{d,f } in comparison to \participants~moving in the opposite direction (Methods).
Activity levels change significantly immediately after relocation and are symmetric but inverted for \participants~moving in the opposite direction.
All error bars throughout this paper correspond to bootstrapped 95\% confidence intervals.
}
\label{fig:panel1}
\end{figure}

\begin{figure}[tb]
\vspace{-15mm}
\centering
\includegraphics[width=\textwidth]{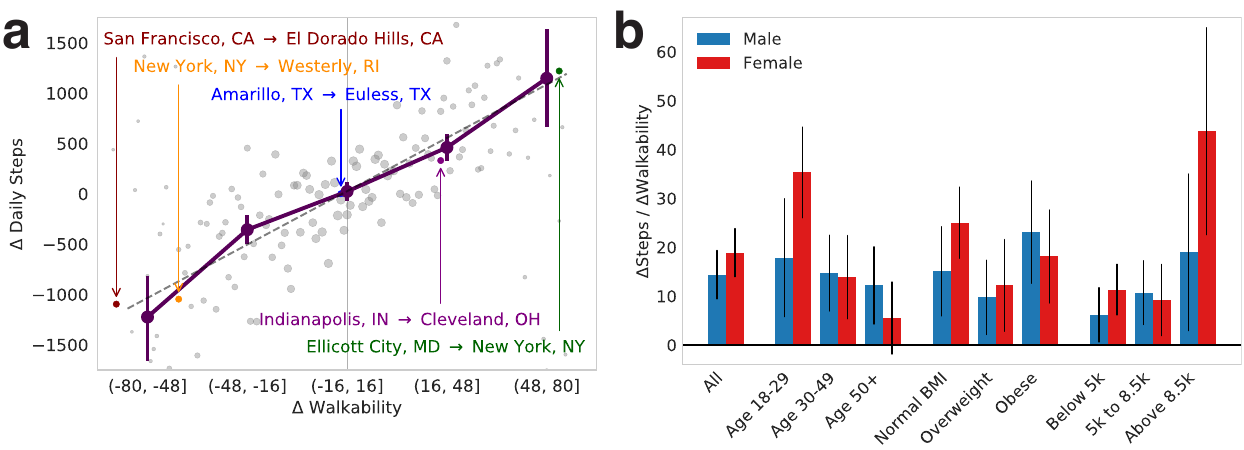} 
\vspace{-10mm}
\caption{
\textbf{Relocations with changes in walkability are associated with corresponding changes in physical activity across most demographics.}
\textbf{a, }
Difference in average daily steps aggregated across all relocations.
We find that significantly more walkable locations are associated with increases of about 1100 daily steps, and significantly less walkable locations are associated with similar decreases (for 49-80 point Walk Score increase/decrease). %
Moving to locations of similar walkability is associated with unchanged physical activity levels.  
\textbf{b, }
Higher walkability is associated with increased daily steps across age, gender, BMI, and baseline activity level groups. 
Bars show the steps gained per day for each point increase in walkability score (assuming linear model; Methods). 
Positive values across all bars reveal that, with increasing walkability, more steps are taken by every subgroup, which is significant for all the subgroups except women over age 50 (Student's t-tests, all $P < 0.05$; women over 50 $P = 0.14$). 
}
\label{fig:panel2}
\end{figure}

\begin{figure}[tb]
\vspace{-25mm}
\centering
\includegraphics[width=.95\textwidth]{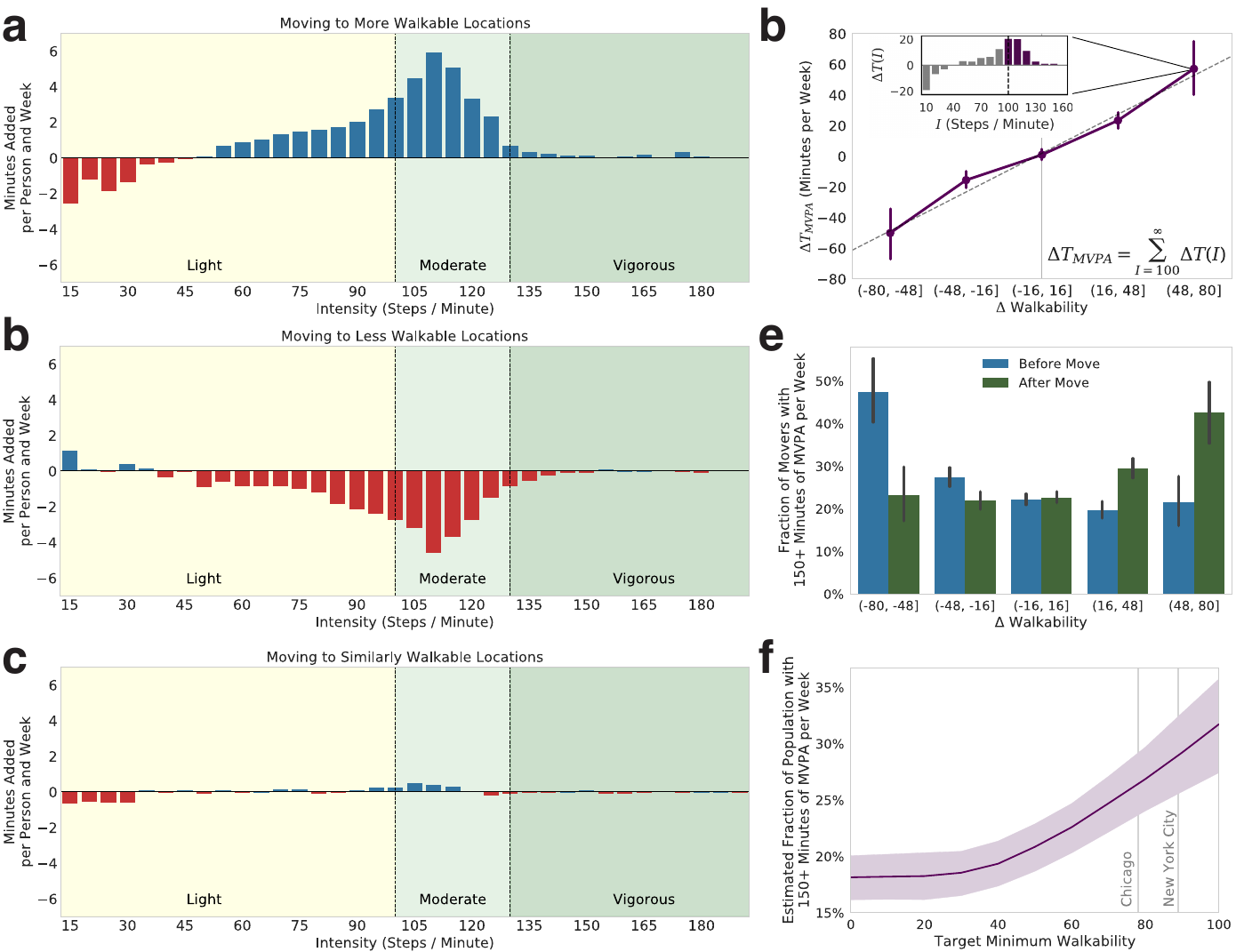} 
\vspace{-4mm}
\caption{
\textbf{Improvements in walkability are associated with increases in moderate-to-vigorous physical activity (MVPA) and with twice as many \participants~meeting aerobic physical activity guidelines (49+ point increase).}
\textbf{a-c,} Changes in physical activity stratified by intensity of physical activity (steps/minute) following relocation to more (a; more than 16 point walkability increase), less (b; more than 16 point walkability decrease) and similarly walkable environments (c; 16 point walkability difference or less).  
\textbf{a,}
We find that walkability-induced additional physical activity (Figure~\ref{fig:panel2}a) predominantly consists of MVPA, which has been shown to be beneficial for many health outcomes.\cite{Lee2012,WHO2010} %
\textbf{b,} 
Moving to less walkable locations is associated with a symmetric loss of MVPA that is equivalent to the increase in more walkable locations (a). 
\textbf{c,}
Further, moving to similarly walkable locations is associated with an unchanged distribution of intensity levels.
This suggests that relocation, in and of itself, is not generally associated with increases in physical activity, for instance due to an individual's motivation to increase physical activity. 
\textbf{d,}
Change in MVPA (minutes/week) versus differences in walkability. 
$\Delta T(I)$ is defined as the change in weekly minutes of activity at intensity level $I$ after relocation, in units of steps per minute.
$\Delta T_{MVPA}$ is computed by summing $\Delta T(I)$ for $I \geq 100$ (inset).
Large increases in walkability (\ie, 49-80 points) are associated with an increase of about one hour per week in MVPA.
\textbf{e,}
The increases in time spent in MVPA lead to twice as many \participants~meeting national and international aerobic physical activity guidelines of 150 min/week or more in MVPA (before \pctMVPAbefore, after \pctMVPAafter).
\textbf{f,}
A simulation based on these estimates predicts that if all U.S. cities had the walkability of Chicago or Philadelphia (Walkability score 78), subjects would increase their average activity by 443 more daily steps and 24 more minutes of MVPA per week, and \pctchangeCHIPHI or \millionmoreCHIPHI million additional Americans would then meet national physical activity guidelines for MVPA (Methods).
}
\label{fig:panel3}

\end{figure}

\clearpage

\section{{\Large Main Text}}

\ifheaders
\xhdr{P1: What are the key questions and limitations that we want to address?}
\fi
A substantial number of people worldwide are physically inactive\cite{Hallal2012,sallis2016progress,kohl2012pandemic} and therefore at risk for common and deadly noncommunicable diseases such as cardiovascular disease, cancer and dia\-betes\cite{WHO2010,Lee2012,sallis2016progress}. 
Meanwhile, urban environments worldwide have grown rapidly with current estimates predicting 6.7 billion people living in cities by 2050\cite{un2018worldurbanization}.
While the evidence base on the impacts of the design of urban environments on physical activity levels has grown, further information is needed on the putative causal impacts of diverse urban environments on key health behaviors such as physical activity\cite{deweerdt2016urban,Sallis2016,prince2022,Ding2018,stearns2023} and interactions between environmental and individual factors\cite{Bauman2012}. 
Specifically, current evidence has not been able to disentangle whether physical activity levels are directly influenced by the built environment or are mainly a product of personal preferences\cite{mccormack2011search, Ding2018}.
Understanding these factors is critical for developing optimal public policy\cite{dhhs2008_activityguidelines,Chokshi2014,graham2013designs} as well as for planning cities\cite{Sallis2016,bettencourt2010unified} and designing behavior change interventions\cite{Servick2015,reis2016scaling}.

\ifheaders
\xhdr{P2: Limitations of prior work}
\fi
Previous studies on the effects of the built environment on physical activity have led to mixed or modest findings and have not been able to reliably distinguish between direct environmental impacts and individual preferences. 
Common methodological limitations include small cohorts, single or only a few locations, over-reliance on self-reported activity, with its attendant biases\cite{Prince2008}, cross-sectional designs that constrain temporal understanding and causal inference, residential self-selection, and other confounding factors\cite{deweerdt2016urban,mccormack2011search,goenka2016our,giles2013influenceRESIDE,Sallis2016,ludwig2012neighborhood}. 
Today's mobile phones, including the now globally dominant smartphone, can capture physical activity and geolocation in a continuous fashion and are a powerful tool for studying large-scale population dynamics and health\cite{Servick2015}, revealing the basic patterns of physical activity\cite{althoff2017large}, sleep\cite{Walch2016}, human movement\cite{Gonzalez2008}, and mood rhythms\cite{golder2011diurnal}, along with the dynamics of the spread of diseases such as malaria\cite{Wesolowski2012} and COVID-19\cite{chang2021mobility}, and linkages with socioeconomic status in developing countries\cite{blumenstock2015predicting}.
In this study, we use a large-scale physical activity dataset to disentangle built environment influences from personal proclivities through a natural experiment, and quantify the impact of walkability, on changes in individual and population physical activity levels.

\ifheaders
\xhdr{P3: Talk about our dataset and study.}
\fi
We study \numdaysThirty days of minute-by-minute step recordings from \numusers~users of the Azumio Argus smartphone application, who relocated at least once within a 3-year observation period (Supplementary Table~\ref{table:table_demographics}).
Overall, these \participants~relocated a total of \nummoves times among \numcities cities within the United States, forming a nationwide natural experiment (Figure~\ref{fig:panel1}a). 
The dataset includes smartphone-derived accelerometry recordings of physical activity
for free-living individuals that were exposed to different built environments, enabling us to compare their objectively measured, longitudinal physical activity for up to 90 days pre- and post-relocation (Figure~\ref{fig:panel1}b). 
The average \participant~recorded 5574 steps per day
(standard deviation $\sigma$ = 3055) over an average span of 14.2 hours. 
Research has demonstrated that smartphones provide accurate step counts\cite{case2015accuracy} and reliable activity estimates in both laboratory and free-living settings\cite{hekler2015validation}. 
Previous work further verified that data from the Argus smartphone application reproduced established relationships between age, gender, weight status, and activity, as well as country-level variations in activity and obesity levels\cite{althoff2017large}.

\ifheaders
\xhdr{P4: NYC example, new environment}
\fi
Our large-scale activity measurements enable us to characterize the impact of built environments on physical activity. 
Consider the 178 \participants~relocating to New York City (walkability 89/100) coming from various less walkable U.S. locations (Figure~\ref{fig:panel1}c; at least one standard deviation or 15.4 points lower Walk Score; mean walkability 48). 
When exposed to the built environment of New York City after relocating, these \participants~increased their physical activity by 1,400 steps from 5,600 to 7,000 average daily steps ($P < 10^{-10}$, Figure~\ref{fig:panel1}e).
\Participants~relocating in the opposite direction, i.e., from New York City to other less walkable U.S. cities (Figure~\ref{fig:panel1}d), exhibited an inverted, symmetric effect of decreasing their physical activity by 1,400 steps from 7,000 to 5,600 average daily steps ($P < 10^{-10}$, Figure~\ref{fig:panel1}f; more examples in Figure~\ref{fig:extra_moving_examples}).

\ifheaders
\xhdr{P5: Fig2a, aggregating, generalization, confounders}
\fi
To investigate whether moving to more walkable environments generally leads to increased physical activity, we aggregate changes in physical activity across all relocations in the dataset (Figure~\ref{fig:panel2}a; Methods). 
We find that relocations to more walkable cities (Walk Score increases of 49 and higher) are associated with increases of about 1100 daily steps, equivalent to 11 minutes of additional walking activity every day\cite{marshall2009translating}. 
While individuals may be more motivated to walk more immediately after relocation, we observe that these increases are sustained  over three months after moving (Figure~\ref{fig:why_month_average}c and Figure~\ref{fig:180_day_effects}). 

Of note, individuals moving to environments with walkability scores similar to the environment from which they came exhibited unchanged activity levels, while the physical activity increases observed with moves to more walkable environments  mirrored the physical activity decreases observed when moving to less walkable environments (Figure~\ref{fig:panel2}a; Figure~\ref{fig:si_move_heatmap_quintiles}). 
Thus, the relationship between walkability and daily steps appears to be monotonic and symmetric. 
We find similar, consistent effects of walkability increases and decreases (Methods) between cities in similar climates (\eg, Ellicott City, MD to New York, NY in Figure~\ref{fig:panel2}a and more generally across relocations during all seasons (Figure~\ref{fig:si_main_effect_by_season}), and after relocating to cities of higher, similar, and lower median household income (Figure~\ref{fig:si_main_effect_by_income}).
These results suggests that physical activity levels are directly influenced by the built environment and not simply a product of personal preferences or other types of confounding in the dataset.

\ifheaders
\xhdr{P6: 2b: does everyone benefit?}
\fi
We find that higher walkability is associated with significantly more daily steps across all age, gender, body mass index (BMI), and baseline activity level groups, %
which is significant for all the subgroups except women over age 50
(Figure~\ref{fig:panel2}b; Student's t-tests, all $P < 0.05$; women over age 50, $P = 0.14$). 
\disc{Previous research has identified additional barriers to physical activity relevant to older women including cultural expectations, norms, societal messages discouraging physical activity, family priorities, and safety\cite{vrazel2008overview,eyler2002environmental}}. 
The relationship between walkability and activity is strongest for highly active women (gaining 43.7 steps per walkability point increase).
Importantly, we find that regardless of BMI status, individuals record more steps after moving to more walkable cities, and that these increases are also shared by individuals who were less active prior to moving (Figure~\ref{fig:panel2}b). 
\disc{These findings suggest that compared to interventions targeting individuals and reaching small numbers of people, changes to the built environment can influence large populations.
However, the relatively smaller effect for older women suggests that, for this group in particular, built environment changes may need to be accompanied by additional age- and gender-specific interventions aimed at their specific constraints.}

\ifheaders
\xhdr{P7: What kind of activity is added?}
\fi
Next, we investigated whether the walkability-induced increase in steps reflected an increase in moderate-to-vigorous physical activity (MVPA), which has been shown to be beneficial for many health outcomes, including lower all-cause-mortality risk\cite{Lee2012,WHO2010}. %
Using minute-by-minute step data, we find that additional steps taken after moving to a more walkable location are predominantly composed of MVPA corresponding to brisk walks (Figure~\ref{fig:panel3}a). 
We estimate that large increases in walkability (\ie, 49-80 points) are associated with an increase in MVPA of about one hour per week (Figure~\ref{fig:panel3}d)
Further emphasizing the consistency and symmetry of built environment effects, we find that similar amounts of MVPA are lost when relocating to a less walkable location (Figure~\ref{fig:panel3}b), and that the activity intensity distribution remains effectively unchanged when relocating to a similarly walkable location (Figure~\ref{fig:panel3}c). 
U.S. national physical activity guidelines recommend, similar to international guidelines, 150 or more minutes per week of MVPA in order to obtain optimal health benefits\cite{PhysicalActivityGuidelinesforAmericans}. 
For a walkability increase between 48 and 80 points increase, we find that the associated increases in MVPA would support \pctMVPAafter~of \participants~meeting guidelines for MVPA versus \pctMVPAbefore~before relocation, a 98\% relative increase (Methods). 
\disc{
Our findings substantively expand previous literature indicating that improving the walkability of built environments can lead to better health outcomes across large populations. }

\ifheaders
\xhdr{P8: how much MVPA is added in total? guidelines}
\fi

\ifheaders
\xhdr{P9: simulation experiment}
\fi
We performed a simulation study to predict how improving walkability would support increasing the fraction of the U.S. population that meets aerobic physical activity guidelines (Methods).
Our dataset covers \numcities U.S. cities, which are home to more than 41\% of the country's population (137 million), and we adjust for age-differences between the smartphone user population and the U.S. adult population (Methods). 
According to our smartphone-based objective measurements, about \pctnowmeetguidelines or \millionnowmeetguidelines million Americans met the guidelines for MVPA between 2013--2016. 
Our estimate of \pctnowmeetguidelines meeting aerobic guidelines is within expectations, given well-established differences between accelerometer-derived and self-reported physical activity (Methods).\cite{prince2008comparison,tucker2011physical}
Our simulation (Figure~\ref{fig:panel3}f) predicts that bringing all U.S. locations to the level of Chicago or Philadelphia (walkability score 78) may lead to \pctchangeCHIPHI or \millionmoreCHIPHI million more Americans meeting aerobic physical activity guidelines. 
Bringing all U.S. locations to the level of New York City (walkability score 89) may lead to \pctchangeNYC or \millionmoreNYC million more Americans meeting these guidelines.

\ifheaders
\xhdr{P10: Limitations and connection to literature, novelty}
\fi
There are limitations to the device-based instrument (i.e., people's personal smartphones) we used to collect physical activity data in their natural environments. 
For example, our sample may be biased towards individuals of higher socioeconomic status and people interested in their activity and health. 
However, we find that walkability improvements led to increased physical activity after relocating to cities of higher, similar, and lower median household income (Figure~\ref{fig:si_main_effect_by_income}).
We further acknowledge that other city characteristics may affect walking and be correlated with the city's walkability (\eg, length of work days). %
However, we find that walkability differences are associated with physical activity differences in cities of similar climate (Figure~\ref{fig:si_main_effect_by_koppen}) and across all seasons (Figure~\ref{fig:si_main_effect_by_season}).
While relocation uniquely enables the quasi-experimental study of behavioral changes in different environments, there may be selection effects driving relocation referred to as residential self-selection\cite{mccormack2011search}. 
However, since we observe no change in activity levels for individuals who moved to a similarly walkable city, and symmetric activity losses in individuals who moved to less walkable cities, this suggests that the changes we observe are not simply due to individuals who move being more motivated to be more active
(Figure~\ref{fig:panel2}a, Figure~\ref{fig:panel3}c-e, Figure~\ref{fig:si_move_heatmap_quintiles}). %
Further, we find that \participants~relocating to locations of higher, similar, and lower walkability are similar in age, weight status, and baseline activity levels, limiting our concerns for potential selection effects based on relocation preferences (Figure~\ref{fig:si_movers_vs_ws_diff}). 
In addition, the population of relocating \participants~closely matched U.S. Census estimates in terms of age and gender distribution (Figure~\ref{fig:si_azumio_vs_census}; Methods). 

Over 90\% of adults in the United States already own a smartphone\cite{Pew2024smartphones} and the number of mobile connections worldwide has risen to 8.5 billion\cite{EricssonMobilityReport2023}, exhibiting significant year-to-year increases.
Therefore, we expect any biases related to smartphone ownership and usage to continue to diminish in the future. 
This study is restricted to a single country and results may not generalize to other countries. 
However, previous studies have found, in general, similar types of built environment relationships across countries diverse in climate, demographics, income, culture, and activity supportiveness\cite{Sallis2016,kerr2013advancing,adams2014international,cerin2014neighborhood}.
Since these studies employed walkability indices that were based on elements shared with the measure used here (Methods), this suggests that our findings may generalize to other countries.
We chose a simple, highly used, and extensively validated measure of walkability at a city level\cite{Duncan2011,carr2010walk,carr2011validation,manaugh2011validating,hirsch2013walk}. 
However, this type of aggregated, non-divisible walkability score precludes the ability to identify which elements of walkability may confer the largest benefits. 
Further research is indicated to identify key environmental features on a neighborhood level and disentangle their individual contributions, building on past cross-sectional research and smaller-scale studies using self-report physical activity measures which currently constitute the majority of research in the field\cite{smith2017systematic}. %
While walking is the most popular aerobic physical activity\cite{cdc2012walking}, our dataset may fail to capture time spent in activities where it is impractical to carry a phone (\eg, soccer) or steps are not a major component of the activity (\eg, bicycling), and there may exist systematic differences in wear time, because \participants~in the current dataset had to carry their phone for steps to be recorded. The increasing prevalence of wearable activity trackers in the form of smartwatches and similar devices will continue to enable more convenient methods of capturing daily movement and steps.
However, our smartphone dataset reproduces previously established relationships between activity across geographic locations, gender and age\cite{althoff2017large}.
Further, we find that the span of time over which steps were recorded is uncorrelated with relocating to higher or lower walkability areas (Figure~\ref{fig:si_weartime}), and thus systematic wear time differences are unlikely to affect our analyses. 
Together, these results increase confidence that our 
dataset is able to identify activity differences between built environments and groups based on gender, age, and weight status.

\ifheaders
\xhdr{P11: Summarize findings and implications, novelty, study strengths}
\fi
This countrywide natural experiment presents prospective evidence of built environments affecting physical activity across \nummoves relocations among \numcities U.S. cities over a 3-year timespan. 
It reveals the direct behavioral impacts of differing built environments on individuals' physical activity levels 
and demonstrates the utility of such massive digitally-enabled real-world datasets for evidence-based policy.
Our findings suggest that designing built environments to be more activity-friendly could have significant effects on the physical activity of large populations, and serve as a powerful complement to interventions that focus on changing behavior at the individual-level. %
However, changes in built environments may need to be accompanied by additional age- and gender-specific interventions aimed at specific subgroups who could particularly benefit from physical activity increases (i.e., women over 50 years old). 
The quality of the prospective device-collected evidence and consistency of findings across numerous cities, demographic groups, and relocation-related walkability differentials highlight the fundamental importance of the urban built environment in improving physical activity and health.

\clearpage
\section{{\Large Methods}}

\subsection{Study Design}
We conducted a countrywide, prospective, longitudinal physical activity study of United States residents that evaluated their physical activity levels within the context of the walkability of their built environments before and after relocation ("\participants"). 
We leveraged the naturally occurring physical activity data that was captured by a health app on \participants' phones to compare each person's physical activity levels before and after they relocated to a different U.S. area. 
While similar relocation-based study designs have been used previously to estimate effects of place and built environments\cite{giles2013influenceRESIDE,chetty2018impacts,christie2022cross}, the vast majority have been limited by relatively small sample sizes, use only self-report physical activity measurement, and the limited diversity with respect to the areas to which they relocated. 
Objective measures of both urban walkability and physical activity were used and are discussed in more detail below.
We analyzed anonymized, prospectively collected data from \numtotalusers~U.S. smartphone users employing the Azumio Argus health app over three years (March 2013 to February 2016) to identify \numusers~\participants~that relocated \nummoves times among \numcities U.S. cities. 
These \numcities cities are home to 137 million Americans, or more than 42\% of the United States population.
We note that relocation is neither purely exogenous nor random and discuss important implications below.
We follow established best-practices for analyzing large-scale health data from
wearables and smartphone apps\cite{hicks2019best}.

The Azumio Argus app is a free smartphone application for tracking physical activity and other
health behaviors. 
\participants~were excluded from a particular analysis if necessary information was unreported (for example, \participants~with no reported age were excluded from the analysis of Figure~\ref{fig:panel2}b).
Table~\ref{table:table_demographics} includes basic statistics on study population demographics and weight status (Body Mass Index; BMI).  
Data handling and analysis was conducted in accordance with the guidelines of the Stanford University Institutional Review Board.

\subsection{Identifying \Participant~Relocation} 
We defined \participant~relocation as the action of moving to a new place for a substantial amount of time.
We identified \participant~relocation as follows. \Participant~location on a given day was assigned to a city based on the weather update in the \participant’s app activity feed. Weather updates are automatically added to the feed of each \participant~according to the nearest cell phone tower. 
We searched for \participants~that stayed in one location within a 100 km radius for at least 14 days,
and then moved to a different location that was at least 100 km away. \Participants~were required to stay within a 100 km radius of this new location for at least another 14 days.
The 14 day threshold was chosen to filter out short trips that may be related to business or leisure travel. 
Using this threshold, we find that most \participants~do not relocate again and spend a median of 81 days in the new location, effectively excluding the impact of short-term travel on our analyses. Most \participants~stopped to track their activity at this time, rather than relocating again. %
In addition, we repeated our analyses with thresholds of 21 and 30 days and found highly consistent results (Figure~\ref{fig:fig2a_by_move_def}).
We required a substantial move distance (100 km or more) to ensure that relocating \participants~were exposed to a new built environment. 
We allowed for up to 5 days of intermediate travel between these two locations and ignored these days during analyses.
We applied this method to \numtotalusers~users of the Argus smartphone app and identified \internummoves relocations.
Among these, we required \participants~to have used the app to track their physical activity for at least 10 days within 30 days before and after their relocation (as in previous work\cite{althoff2017large}). 
We further required at least one day of tracked physical activity pre- and post-relocation to ensure that whenever we compare two \participant~populations, that these populations are identical and therefore comparable (i.e., we seek to identify within-\participant~changes in physical activity).
We repeated our statistical analyses with alternative data inclusion criteria, such as the number of days with tracked physical activity, and found similar results.

\subsection{Physical Activity Measure} 
Our device-based (historically often called "objective") measure of physical activity was the number of steps over time recorded by the \participant's smartphone.
Steps were determined based on the smartphone accelerometers and the manufacturer's proprietary algorithms for step counting. 
The Azumio Argus app records step measurements on a minute-by-minute basis.
Table~\ref{table:table_pa_stats} includes basic statistics on physical activity and tracking in the study population.

Data from the Azumio Argus app have been used previously to study physical activity in large populations\cite{althoff2017large,althoff2017online,shameli2017gamification}, where the authors showed that this form of data follows well-established trends.\cite{althoff2017large} 
For example, they demonstrated that activity decreased with increasing age\cite{Bauman2012,Hallal2012,Bassett2010,Troiano2008} and BMI\cite{VanDyck2015,Bauman2012,Troiano2008}, and is lower in females than in males\cite{Bauman2012,Hallal2012,TudorLocke2009,Bassett2010,Troiano2008}, trends which are consistent with national surveillance data in this area.
Physical activity estimates were also reasonably well correlated with self-report-based population estimates on a country level\cite{althoff2017large}.

Several studies have established significant differences between accelerometer-derived and self-reported physical activity.\cite{prince2020comparison,tucker2011physical} Self-reports typically overestimate moderate and vigorous activity and underestimate sedentary activity.\cite{prince2020comparison}. 
In a U.S. study using NHANES 2005–2006 data, 59.6\% of adults self-reported meeting MVPA guidelines for aerobic physical activity, while estimates using accelerometry were much lower at 9.6\%.\cite{tucker2011physical} 
For our observation period between 2013-2016, the U.S. National Health Interview Survey reported 49.6-52.6\% meeting MVPA guidelines. Nationally representative accelerometer-based estimates for this time are not available. 
Our smartphone-accelerometry-based estimate of \pctnowmeetguidelines is within expectations, given the differences in measurement and prior data.\cite{tucker2011physical} 
In addition, unlike many prior studies mailing accelerometers to study participants to wear for a week, our study focuses on real-world physical activity by free-living individuals that may not be equally affected by their awareness of being observed (i.e., the Hawthorne effect).

Our estimate of \pctnowmeetguidelines meeting aerobic guidelines is within expectations, given well-established differences between accelerometer-derived and self-reported physical activity (Methods).\cite{colley2018comparison,tucker2011physical}

We filtered out days as invalid when less than 500 or more than 50,000 steps had been recorded. 
We further ignored days immediately preceding and following the relocation itself (5 days before/after relocation), because the process of relocating, rather than the new built environment itself, could impact physical activity during these days.
Physical activity was relatively stable outside this period (Figure~\ref{fig:si_why_remove_mid}). 
We considered physical activity within a window of 30 days before and 30 days after relocation (with the exception of Figure~\ref{fig:why_month_average} and Figure~\ref{fig:180_day_effects} that use 90 day windows to illustrate long-term changes).
In total, our dataset included \numdaysThirty days of objectively-measured minute-by-minute physical activity surrounding \nummoves relocations (\numdaysNinety days for the 180 day period).

We used the following measures as primary outcomes in this study: 
(1) Change in average daily steps following relocation (Figure~\ref{fig:panel1}e-f and Figure~\ref{fig:panel2}a-b).
(2) Change in average weekly minutes spent in moderate-to-vigorous physical activity (MVPA) following relocation,
where we considered all minutes spent at intensities greater or equal to 100 steps/minute as MVPA\cite{marshall2009translating}: 
$\Delta T_{MVPA} = \sum_{I=100}^{\infty} \Delta T (I)$, where $\Delta T(I)$ is defined as the change in weekly minutes of activity at intensity level $I$, in units of steps per minute, after moving.
Figures~\ref{fig:panel3}a-c shows changes in average weekly minutes spent at different intensity levels.
(3) Change in the fraction of the population that met aerobic physical activity guidelines following relocation, 
defined as spending at least 150 minutes per week in MVPA\cite{dhhs2008_activityguidelines} (Figure~\ref{fig:panel3}e-f). 
All error bars correspond to bootstrapped 95\% confidence intervals\cite{efron1994introduction}.

\subsection{Walkability Measure} 
We considered relocations among \numcities cities in the United States.
Walkability scores for these cities were based on the publicly available and systematically developed Walk Score\cite{walkscore2016uscities}. %
Scores are on a scale of 1 to 100 (100 = most walkable) and are based on amenities (\eg, grocery stores, schools, parks, restaurants, and retail) within a 0.25 to 1.5 mile radius (a decay function penalizes more distant amenities) and measures of friendliness to pedestrians, such as city block length and intersection density. 
Table~\ref{table:table_location_walkability} includes basic statistics on the cities included in our study and their walkability scores.

The Walk Score measure is a frequently employed measure of walkability that is freely and widely available across the United States and other countries including Canada and Australia\cite{walkscore2016uscities}. 
It is highly correlated\cite{manaugh2011validating} with other walkability measures\cite{frank2005linking,kuzmyak2006use,porta2005linking}, and was found to offer the best fit to walking trips in a study conducted in Montr\'eal\cite{manaugh2011validating}.
It is widely used in the literature and has been extensively validated\cite{Duncan2011,carr2010walk,carr2011validation,manaugh2011validating,hirsch2013walk,brown2023contributions}.
While other measures of walkability exist\cite{frank2005linking,kuzmyak2006use,porta2005linking}, the Walk Score measure was chosen in light of the pragmatic focus of the investigation and its ease of use and accessibility. 
More comprehensive walkability indices could provide further granular information related to specific aspects of walkability that might be of prime importance. 

Among the \nummoves relocations, 2.4\% (2.4\%) were associated with 49+ walkability point increases (decreases), 20.7\% (21.3\%) were associated with 16-48 walkability point increases (decreases), and 53.1\% of relocations were to similarly walkability locations (-16 to 16 point difference).%

\subsection{Aggregating Relocation-based Quasi-experiments Across United States} 
We aggregated changes in physical activity following relocation based on the difference in walkability scores between the origin and destination city, $\Delta$. 
In Figure~\ref{fig:panel2}a, each circle corresponds to a pair of cities sized by the number of \participants~moving between those cities. 
We fit a linear model  $m \cdot \Delta + b$ to these data with slope $m=16.6$ (Student's t-test; $P < 10^{-10}$) and intercept $b=25.0$ (Student's t-test; $P = 0.462$).

We considered potential confounders such as differences in climate and median income between the origin and destination city. 
We found that the relationship between walkability and walking behavior still holds within pairs of cities with similar climate, for instance moving from Miami, Florida to Jacksonville, Florida, or from Amarillo, Texas to Euless, Texas (see annotations in Figure~\ref{fig:panel2}a as well as more generally in Figure~\ref{fig:si_main_effect_by_koppen}).
Further, we found similar effects across relocations in all seasons (Figure~\ref{fig:si_main_effect_by_season}), and relocations to cities with higher, lower, and similar median household income levels (Figure~\ref{fig:si_main_effect_by_income}).

\subsection{Impact Of Walkability Across Demographics, Weight Status, And Activity Level}
We considered the effect of walkability differences on change in physical activity across subgroups based on demographics, weight status, previous activity level, and gender (ages 18-29, age 30-49, and age 50+ years; normal, overweight, and obese levels of BMI; below 5000, 5000-8500, and above 8500 average daily steps before relocation; males and females).
Due to the approximately linear nature of the relationship between walkability changes and physical activity changes (Figure~\ref{fig:panel2}a), we used a linear model for estimation. 
For each subgroup, we ran independent linear regressions of the difference in daily steps on differences in walkability between cities at the level of individual relocations. 
The models included an intercept coefficient: $m \cdot \Delta + b$. 
We determined the estimated coefficient of walkability ($m$; that is, the increase in daily steps for each one-point increase in walkability of a city) along with 95\% confidence intervals (based on Student's t-distribution) for each subgroup (Figure~\ref{fig:panel2}b). 
We performed Student's t-tests on the regression model coefficients, which establish that relocation to a city of higher walkability is associated with significantly more daily steps across all age, gender, BMI, and activity level groups (Student's t-test; all $P < 0.05$), with the exception of women over 50 years for which the positive difference was not statistically significant (Student's t-test, $P = 0.14$). 
We found that the effect was diminished in overweight and obese women relative to normal-weight women.
Thus, the non-significant effect on women over 50 years of age may be explained in part by the larger average BMI of this group (27.4) compared to other women (25.3; $P<10^{-10}$). %
In comparison, men over 50 years of age also had a larger BMI compared to other men, but the difference was smaller than in women (28.2 vs. 27.0; $P < 10^{-7})$).

\subsection{Adjusting For Seasonality}
Physical activity is influenced by climate and weather\cite{tucker2007effect} and relocations are not equally distributed across seasons (Figure~\ref{fig:si_43_10_90d_month_histogram}). 
We found that differences in physical activity levels following relocations may be influenced by seasonal variation, especially when considering comparatively long observation periods of about six months (Figure~\ref{fig:why_month_average}bc).
For analyses of variation in activity over time (Figure~\ref{fig:panel1}e-f, Figure~\ref{fig:extra_moving_examples}, Figure~\ref{fig:why_month_average}, Figure~\ref{fig:180_day_effects}), we adjusted for these seasonal effects by weighting relocations in each calendar months equally.
This was achieved by first estimating physical activity levels separately for each calendar month and then taking the average.
This process is repeated 1000 times in our bootstrap estimates.

\subsection{Selection Effects In Relocation And Mobile App Usage}
While relocation uniquely enabled the quasi-experimental study of behavioral changes in different environments, there may be selection effects driving relocation, often referred to as residential self-selection.
According to a 2013 U.S. Census Bureau report,
97\% of people moved primarily for reasons of housing, family, and employment\cite{USCensusWhyMove}. 
Less than 1\% of people moved primarily for health reasons. 
In addition, neighborhood selection may be influenced by personal preferences such as exercise and walking activities\cite{mccormack2011search}. 
With respect to this possibility, note that we found no indication of increases in physical activity after moving to a location of similar walkability (Figure~\ref{fig:panel2}a and Figure~\ref{fig:panel3}c). 
This suggests that those relocating \participants~are not simply more motivated to exercise, on average, but that changes in physical activity may be explained by the changing built environment. 
We further acknowledge that other city characteristics may affect walking and be correlated with the city's walkability (\eg, length of work days). %
We investigated potential selection effects further by comparing the population of relocating mobile app users, first, to the overall U.S. population, and second, to the overall mobile app user population, including non-relocating app users. 
We found that the relocating \participant~population is similar in age (36 vs 37.7 median age) and gender (49.8 vs 51.0\% female, $P = 0.132$) to the U.S. population. 
We adjusted for differences in age for the simulation estimates in Figure~\ref{fig:panel3}f and Figure~\ref{fig:si_new_simulation_results}. 
Within the app user population, we found that movers and non-movers (\ie, relocating and non-relocating \participants) tend to be close in age (43.8 vs 37.9 and 38.5 vs 33.7 average age for men and women, respectively; Figures~\ref{fig:si_all_azumio_vs_movers_male_age_hist} and~\ref{fig:si_all_azumio_vs_movers_female_age_hist}), and weight status (68.1 vs 59.8 and 45.6 vs 44.3 percent overweight and obese for men and women, respectively; Figures~\ref{fig:si_all_azumio_vs_movers_male_bmi_hist} and~\ref{fig:si_all_azumio_vs_movers_female_bmi_hist}). 
However, movers were generally more physically active than non-movers (6,284 vs 5,825 and 5,279 vs 4,635 average daily steps for men and women, respectively; Figures~\ref{fig:si_all_azumio_vs_movers_male_PA_hist} and~\ref{fig:si_all_azumio_vs_movers_female_PA_hist}).
Further, we found that within movers, those that relocate to higher, similar, and lower walkability locations were similar in age, weight status, and previous physical activity levels (Figure~\ref{fig:si_movers_vs_ws_diff}).

\subsection{Simulating The Impact Of Walkability Improvements Across The U.S. Population}
We simulated the impact of U.S. nationwide walkability improvements on U.S. population physical activity levels. 
Concretely, we simulated the impact of increasing U.S. city walkability scores to a constant target walkability score between 1 and 100. 
We additionally highlight the walkability scores of Chicago and Philadelphia (78) as well as New York City (89) to aid interpretation.
Since the relocation population was not explicitly drawn to be representative of the U.S. population, we adjusted our estimates through ratio-based post-stratification weights across age-based strata\cite{bethlehem2009applied}. 
We used civilian population estimates from the U.S. Census Bureau for 2016 as the target population distribution. 
While there were no significant differences in the gender distribution~(49.8\% female vs 51.0\% female, $P = 0.132$), 
we found slight differences in age (36.0 vs 37.7 median age) that we corrected for through sampling weights.
We acknowledge that other selection effects and heterogeneous treatment effects may exist. 
Using a bootstrap with 1000 replications, we estimated the difference in the overall U.S. population that would meet U.S. national aerobic physical activity guidelines for MVPA\cite{2018physicalactivityguidelines}
after relocating based on the relocation-induced difference in walkability. 
We used a linear regression model and data from relocations associated with both walkability increases and decreases. 
We estimated the total fraction of U.S. population meeting aerobic physical activity guidelines as the sum between the fraction of people already meeting these guidelines before relocating plus the estimated addition based on the regression model. 
Confidence intervals represent bootstrapped 95\% confidence intervals.
Final estimates are depicted in Figure~\ref{fig:panel3}f and Figure~\ref{fig:si_new_simulation_results}.

\subsection{Role of the Funding Source}
We thank Azumio for donating the data for independent research, and M. Zitnik, E. Pierson, R. Sosic, S. Kumar, and D. Hallac for comments and discussions. 
T.A., B.I., J.L.H, S.L.D., A.C.K., and J.L. were supported by a National Institutes of Health (NIH) grant (U54 EB020405, Mobilize Center, NIH Big Data to Knowledge Center of Excellence). 
T.A. was supported in part by NSF IIS-1901386, NSF CAREER IIS-2142794, Bill \& Melinda Gates Foundation (INV-004841) grants.
J.L.H and S.L.D. were supported by NIH grants P41 EB027060, P2C CHD101913, R01 GM124443, and the Wu Tsai Human Performance Alliance.
A.C.K. King was supported in part by US National Institutes of Health (NIH) National Institute on Aging grant 5R01AG07149002.
Funders were not involved in planning or executing the study and they were not involved in preparing the manuscript. The authors had full access to all of the data in the study and had the final responsibility for the decision to submit for publication.

\xhdr{Data availability} Data are available at\\\url{https://github.com/behavioral-data/movers-public}.

\xhdr{Code availability} Code is available at \\\url{https://github.com/behavioral-data/movers-public}.

\clearpage
\section*{Bibliography}
\bibliography{movers}
\bibliographystyle{naturemag}
\clearpage

\section*{{\Huge Supplementary Information}}

\begin{figure}[tb]
\vspace{-5mm}
\begin{subfigure}{.48\textwidth}
  \centering
  \caption{Moving from New York City, NY.}
  \includegraphics[width=\textwidth]{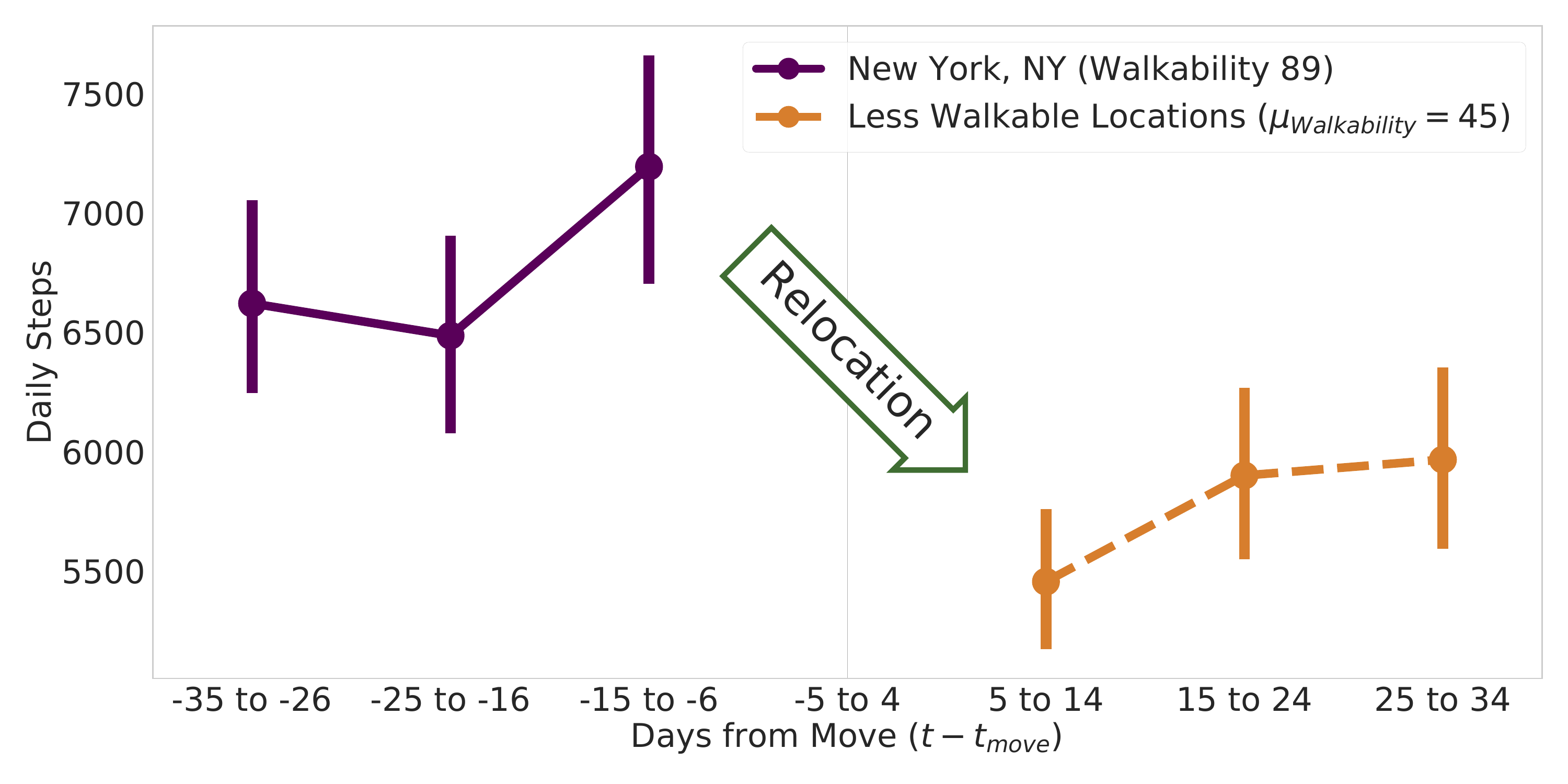}
     \label{fig:si_30d_from_NYC}
\end{subfigure}\hfill
\begin{subfigure}{.48\textwidth}
  \centering
  \caption{Moving to New York City, NY.}
  \includegraphics[width=\textwidth]{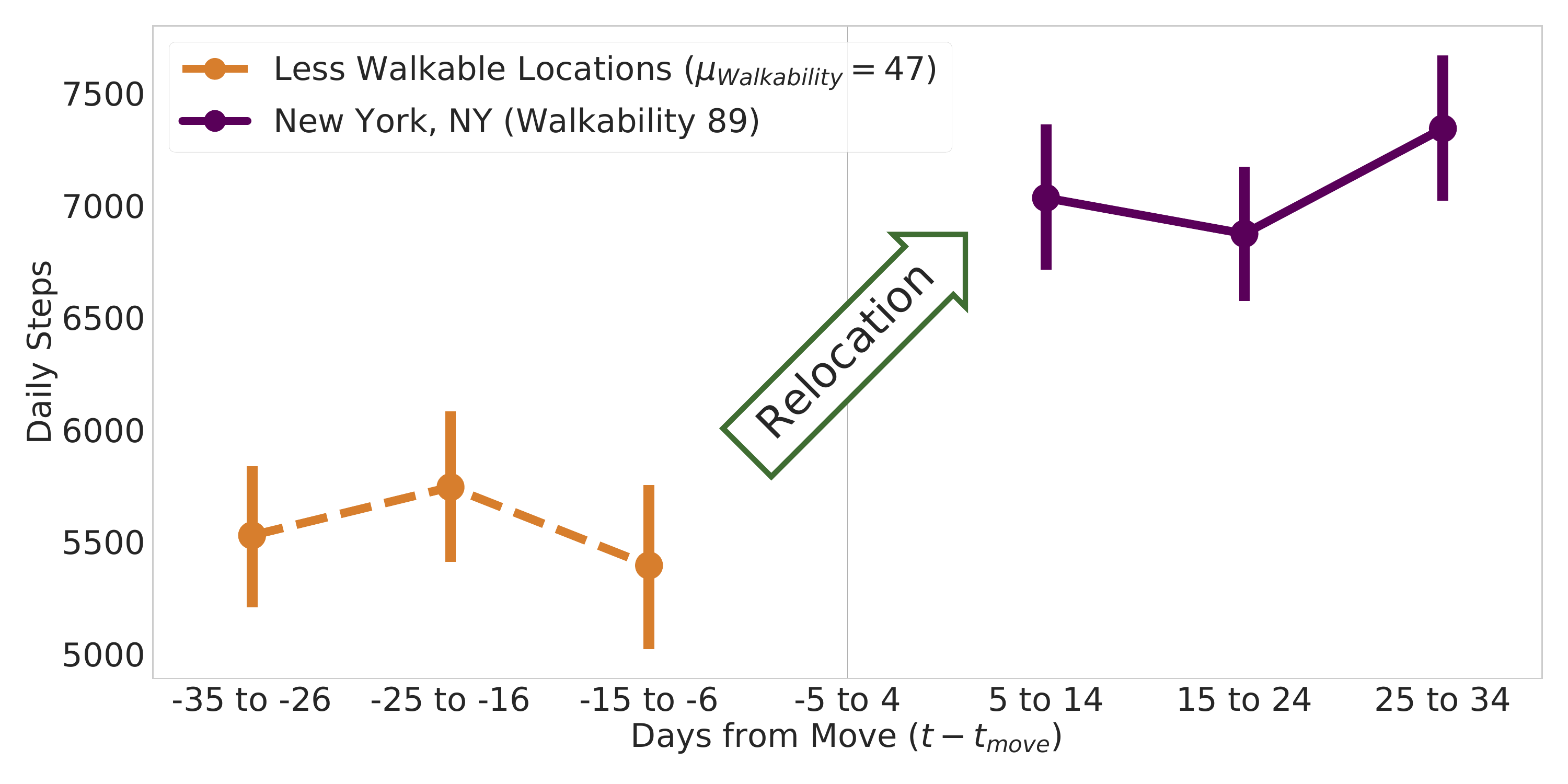}
     \label{fig:si_30d_to_NYC}
\end{subfigure}\\
\begin{subfigure}{.48\textwidth}
  \centering
  \caption{Moving from San Jose, CA.}
  \includegraphics[width=\textwidth]{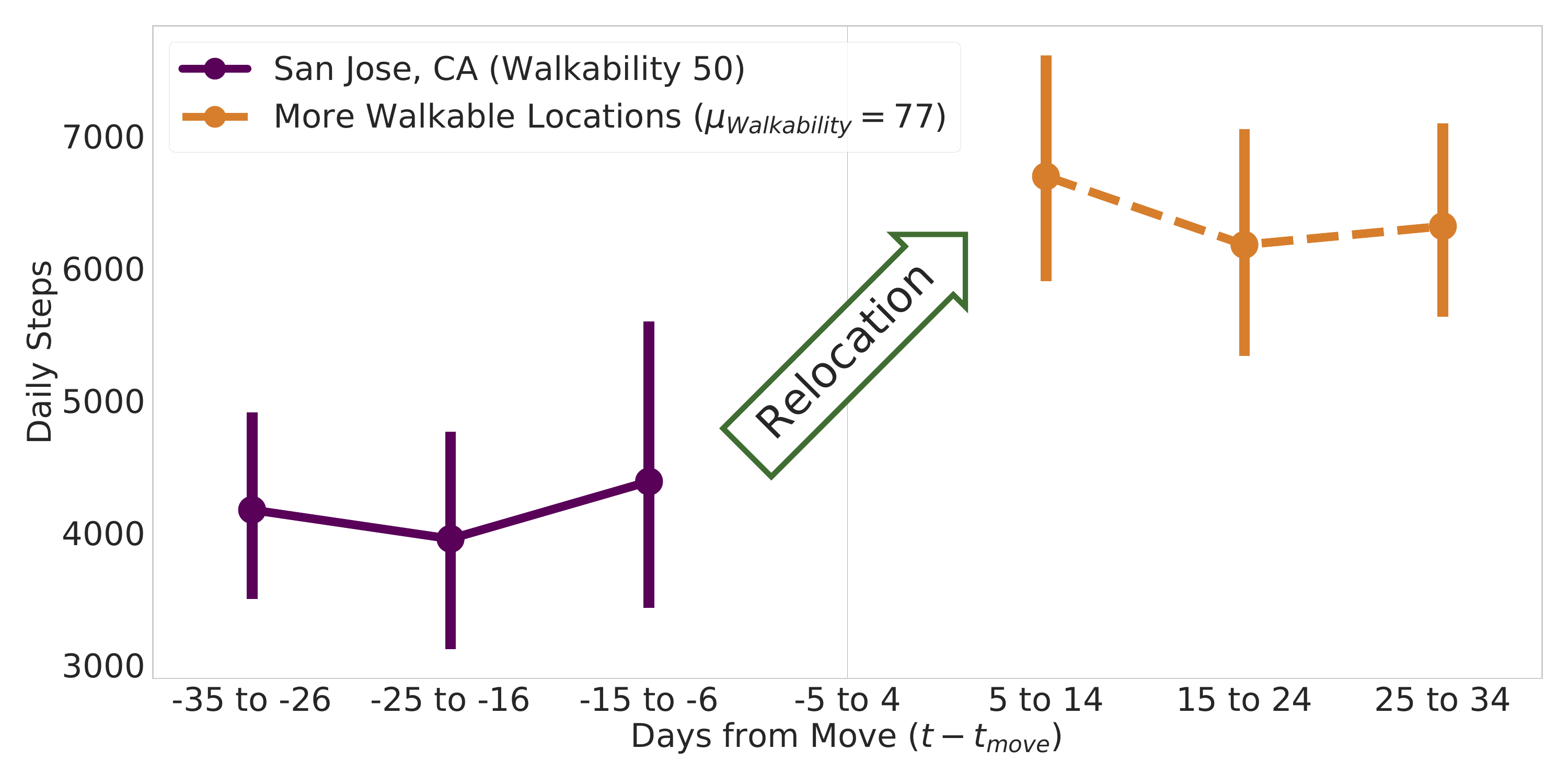}
     \label{fig:si_30d_from_SJC}
\end{subfigure}\hfill
\begin{subfigure}{.48\textwidth}
  \centering
  \caption{Moving to San Jose, CA.}
  \includegraphics[width=\textwidth]{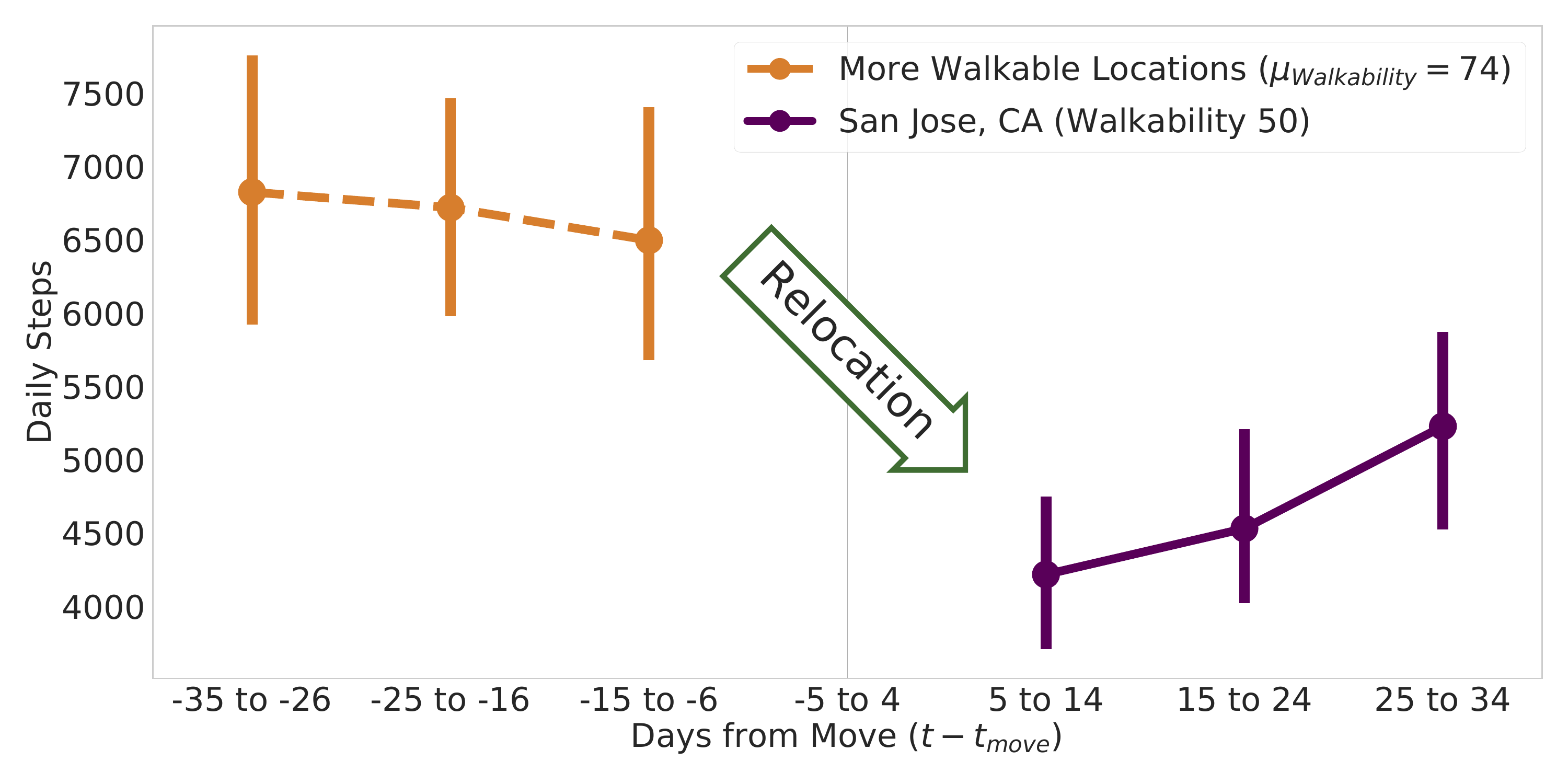}
     \label{fig:si_30d_to_SJC}
\end{subfigure}

\begin{subfigure}{.48\textwidth}
  \centering
  \caption{Moving from Albuquerque, NM.}
  \includegraphics[width=\textwidth]{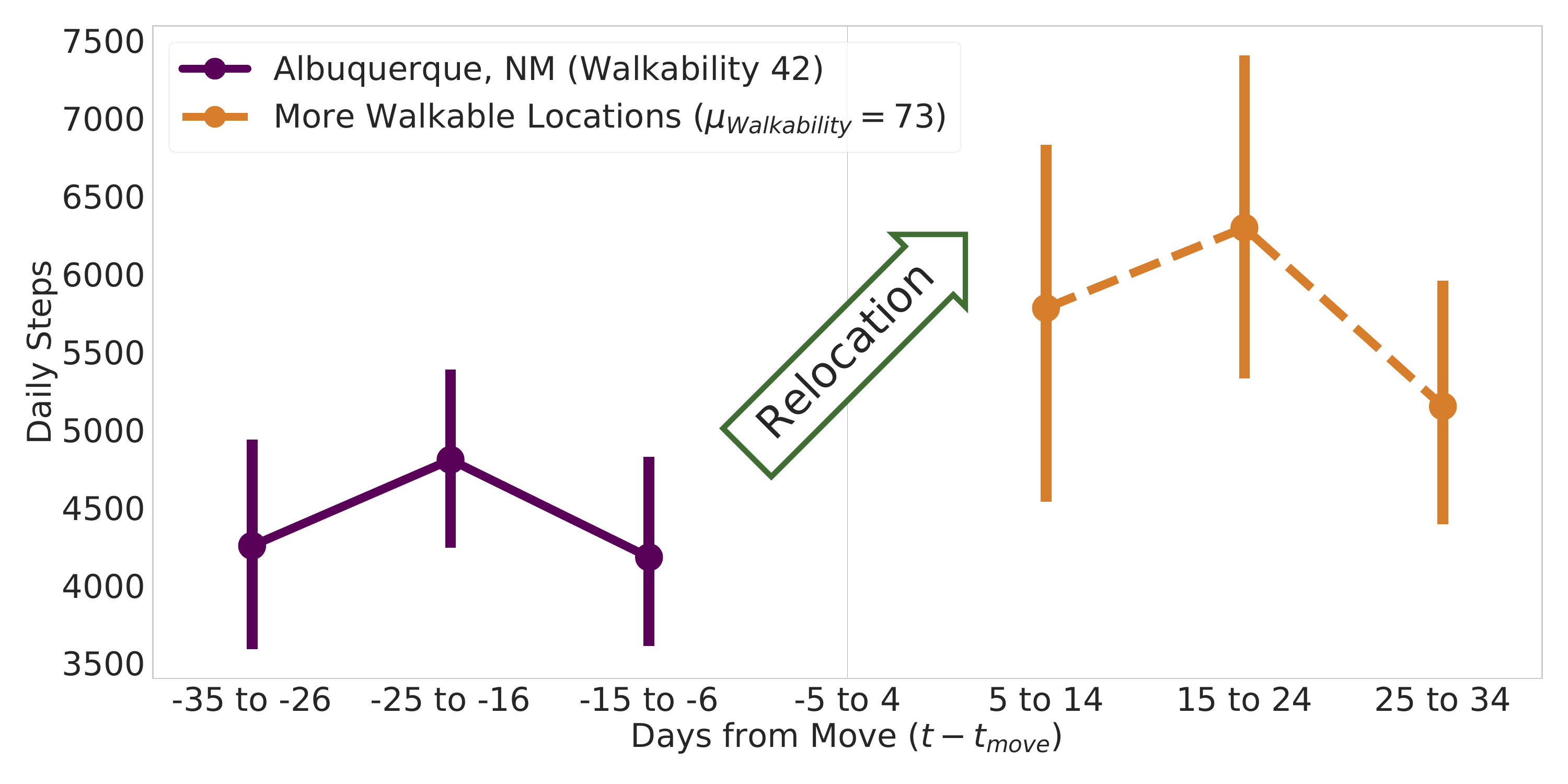}
     \label{fig:si_30d_from_ABQ}
\end{subfigure}\hfill
\begin{subfigure}{.48\textwidth}
  \centering
  \caption{Moving to Albuquerque, NM.}
  \includegraphics[width=\textwidth]{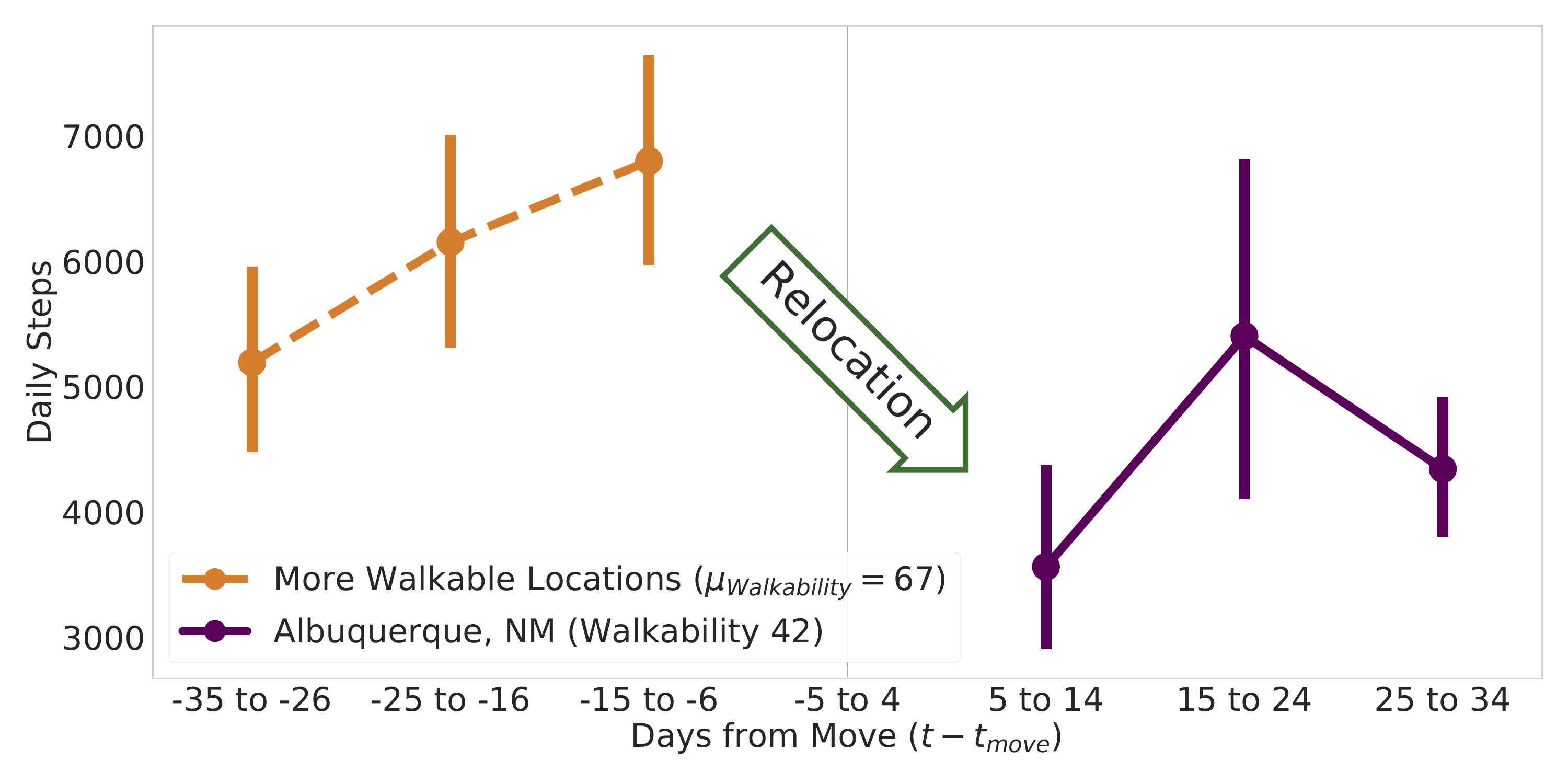}
     \label{fig:si_30d_to_ABQ}
\end{subfigure}\\

\vspace{-10mm}
\caption{
\textbf{\ParticipantsApp~physical activity levels undergo significant changes following relocation to and from specific locations of different walkability.}
Examples show physical activity levels for \participants~moving from/to New York, NY, San Jose, CA, and Albuquerque, NM 
(differences in walkscore of more than one standard deviation of 15.4 points). 
Physical activity levels change significantly by about 1,200 - 1,400 daily steps depending on the location.
Note the symmetry between moving from (left) and to (right) specific locations.
}
\label{fig:extra_moving_examples}
\end{figure}

\begin{figure}[t]
\centering

\begin{subfigure}{0.48\textwidth}
  \centering
  \caption{}
  \includegraphics[width=\textwidth]{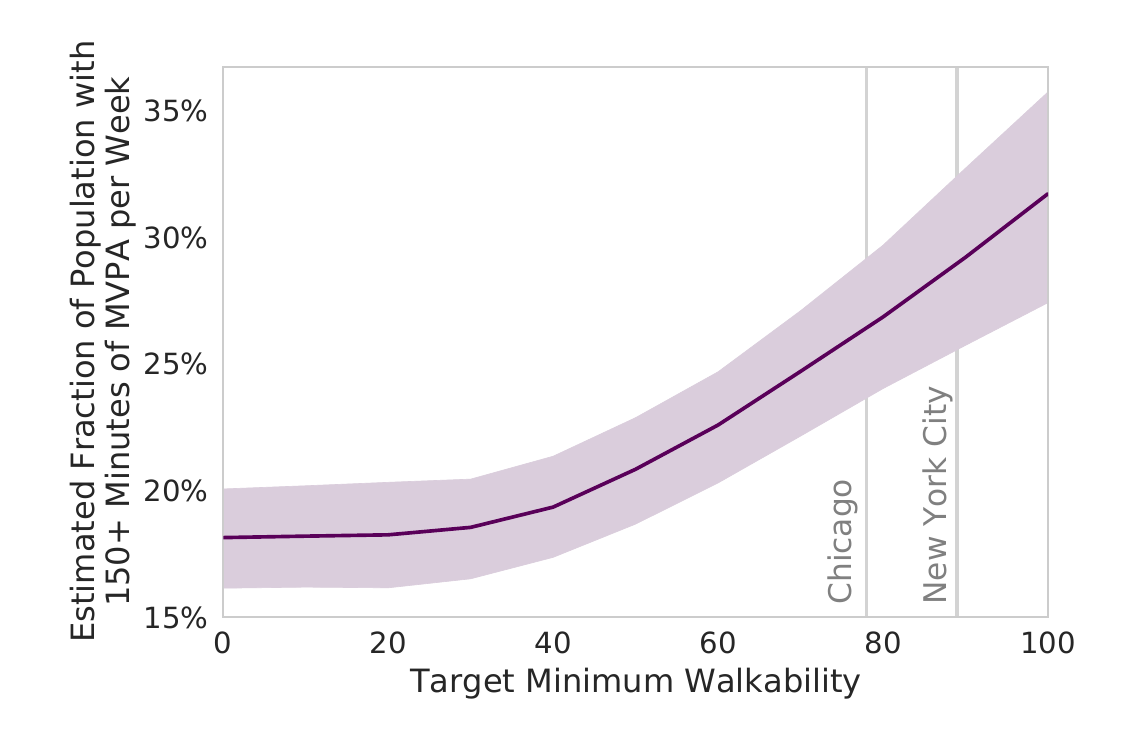}
\label{fig:si_new_fracpop_vs_walkability_obs_weight}
\end{subfigure}\hfill
\begin{subfigure}{0.48\textwidth}
  \centering
  \caption{}
  \includegraphics[width=\textwidth]{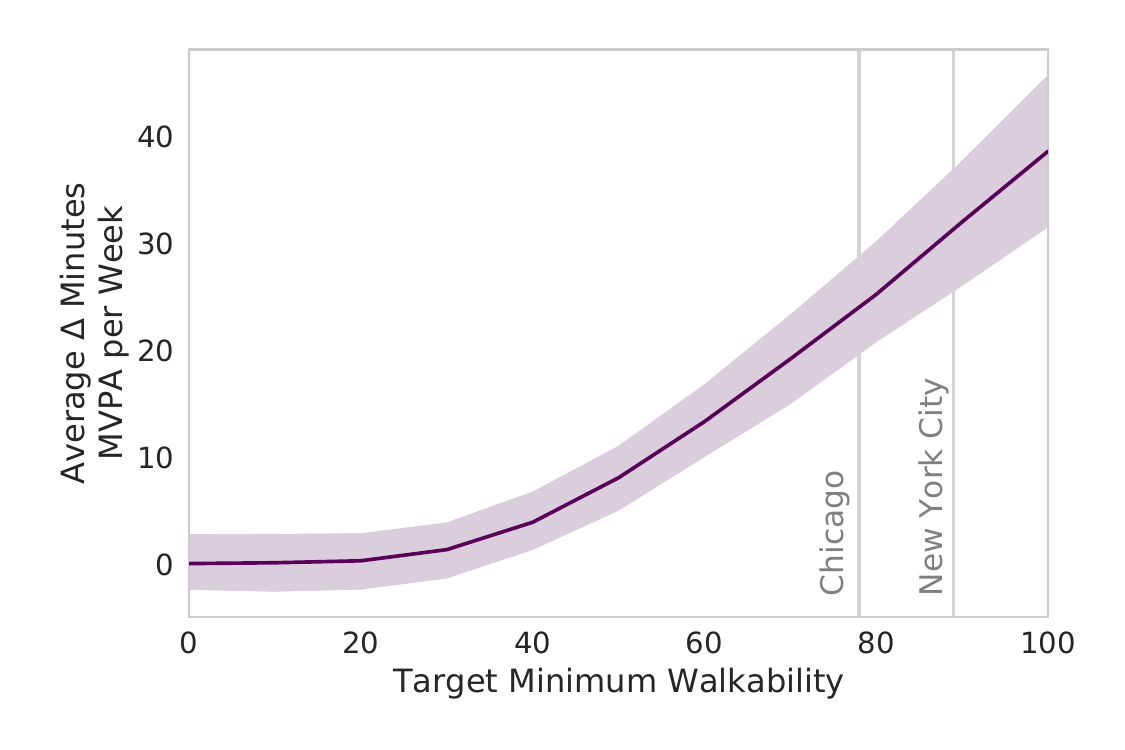}
  \label{fig:si_new_minutes_vs_walkability_obs_weight}
\end{subfigure}

\begin{subfigure}{0.48\textwidth}
  \centering
  \caption{}
  \includegraphics[width=\textwidth]{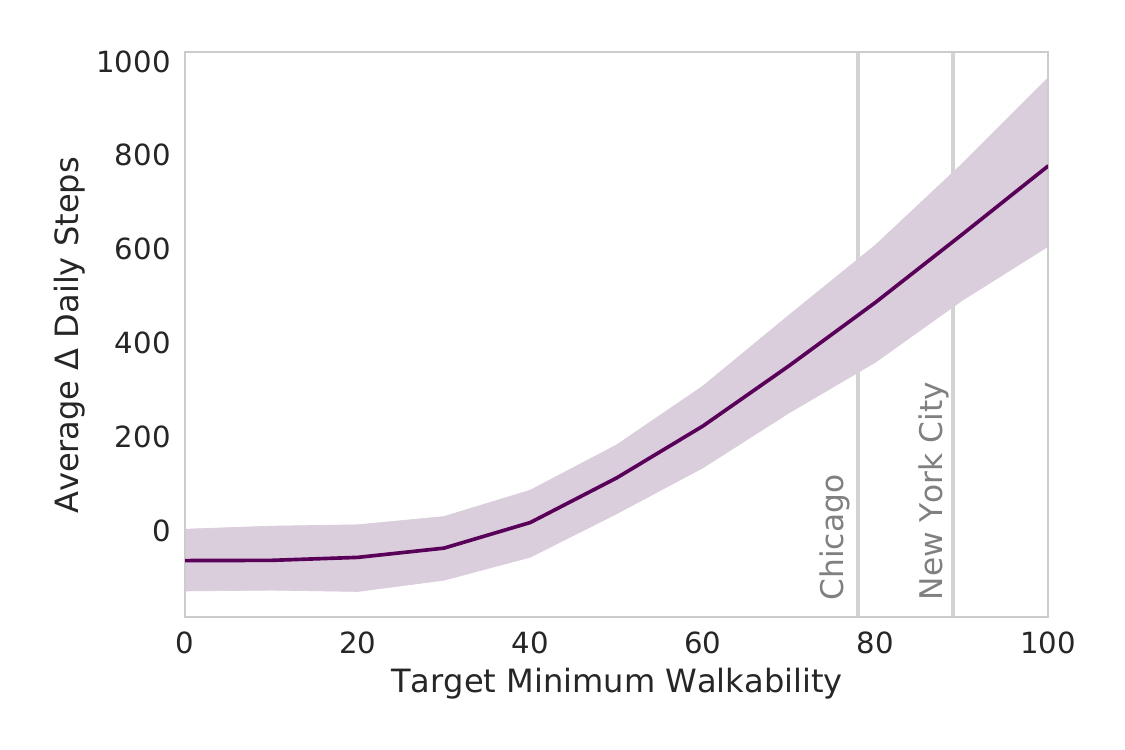}
     \label{fig:si_new_steps_vs_walkability_obs_weight}
\end{subfigure}
\vspace{-5mm}
\caption{
\textbf{Simulating the impact of walkability improvements on United States physical activity levels.}
\textbf{a},
Estimated fraction of population with 150 minutes or MVPA or more per week following an increase in walkability across all represented U.S. locations.
\textbf{b},
Average amount of MVPA added across population following an increase in walkability across all represented U.S. locations.
\textbf{c}, 
Average amount of daily steps added across population following an increase in walkability across all represented U.S. locations.
}
\label{fig:si_new_simulation_results}

\end{figure}

\begin{figure}[t]
    \vspace{-30pt}
    \centering
    {\scriptsize \spacing{1.25}\fontsize{7}{8}\selectfont
        
        \iffigures
        \includegraphics[width=.80\textwidth]{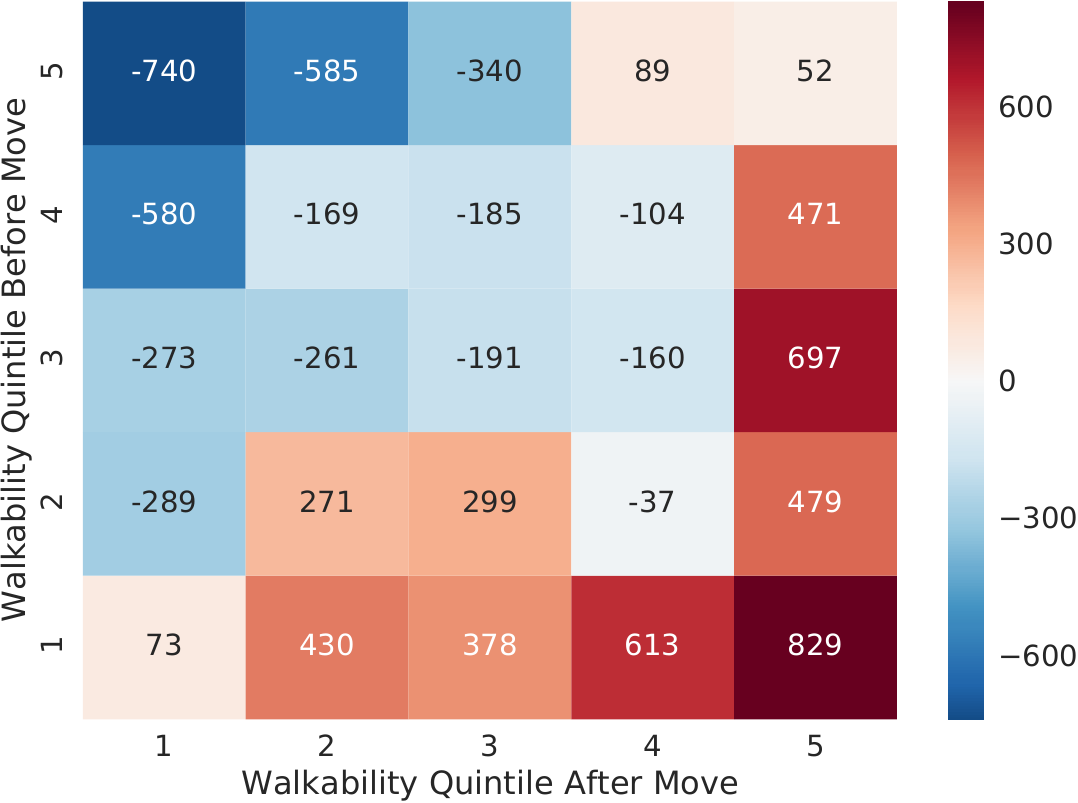} %
        \fi
        
        \caption{%
            \textbf{Changes in average daily steps following relocation between specific walkability score quintiles.}
            Changes in physical activity levels are approximately symmetric and close to zero for relocations to the same walkability score quintile. 
            }
        \label{fig:si_move_heatmap_quintiles}
    } %
\end{figure}

\begin{figure}[t]
\vspace{-5mm}
\begin{subfigure}{\textwidth}
  \centering
  \caption{
  }
  \includegraphics[width=0.5\textwidth]{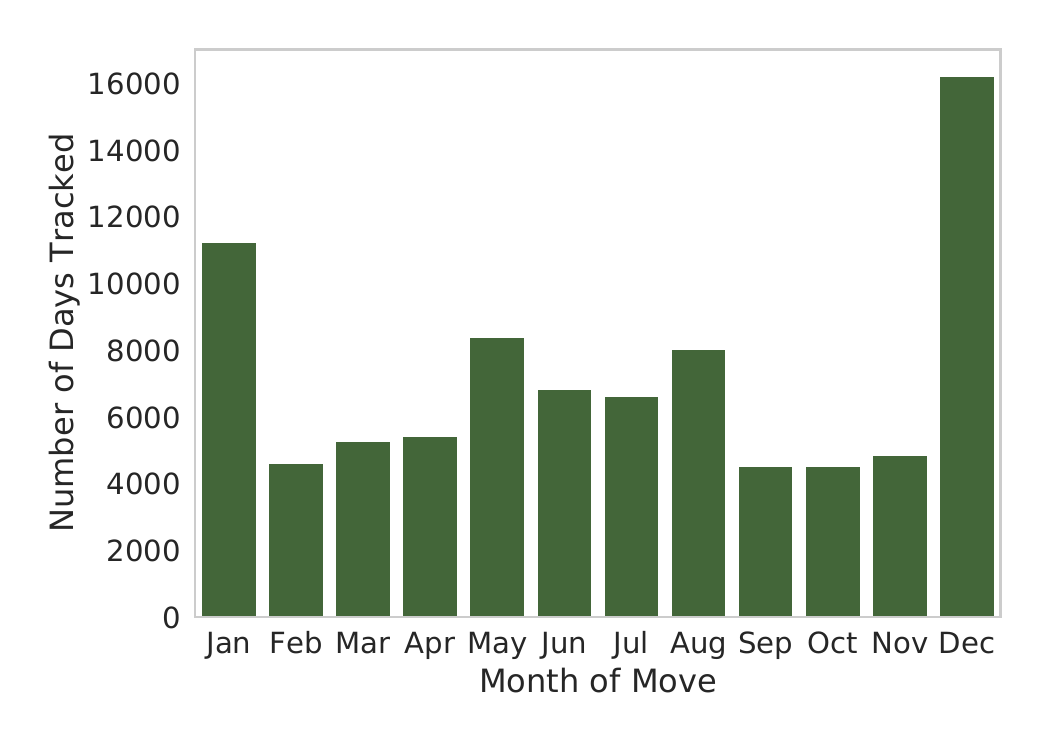}
  \label{fig:si_43_10_90d_month_histogram}
\end{subfigure}
\begin{subfigure}{.5\textwidth}
  \centering
  \caption{
  }
  \includegraphics[width=\textwidth]{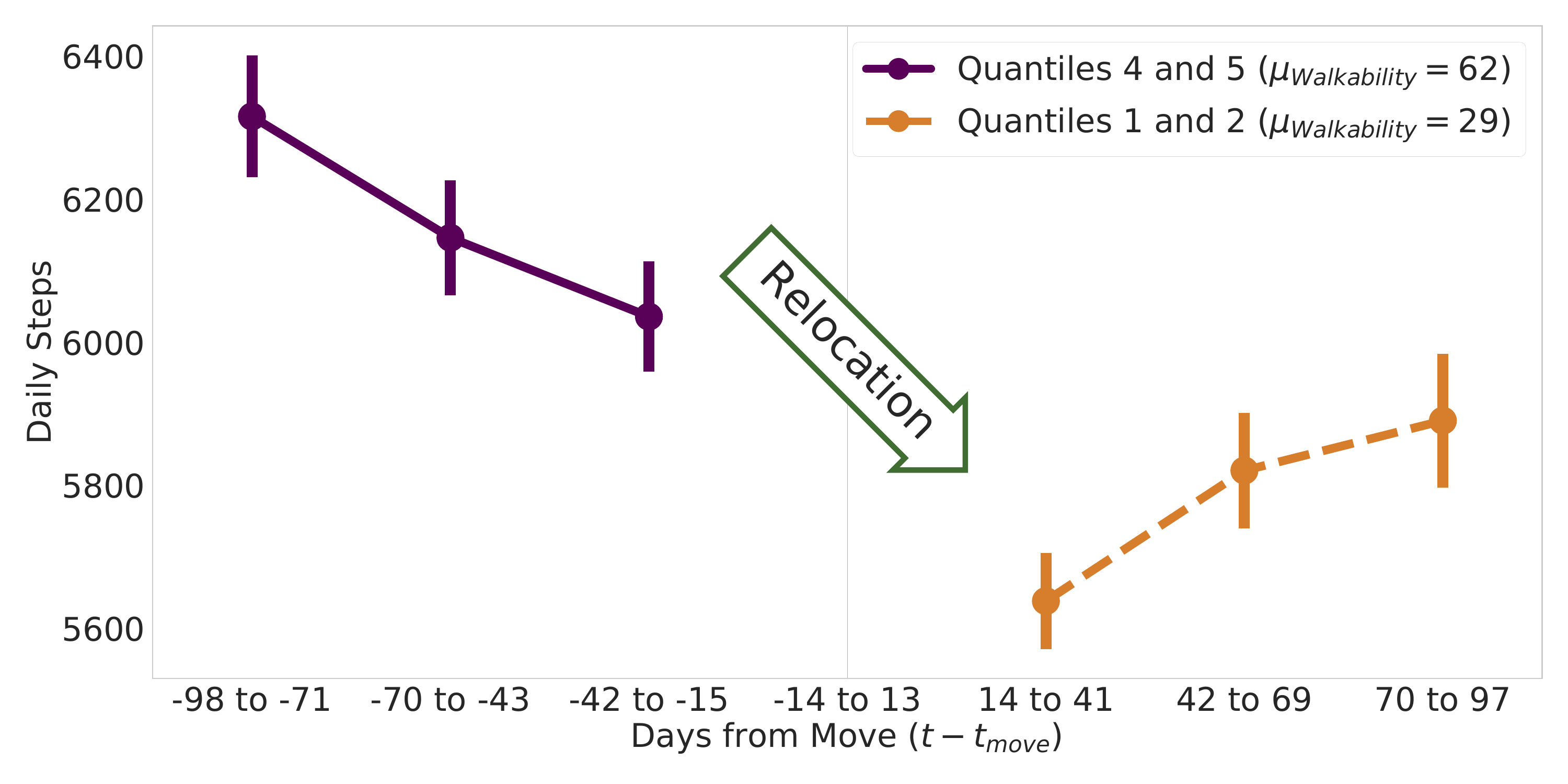}
  \label{fig:si_43_10_90d}
\end{subfigure}
\begin{subfigure}{.5\textwidth}
  \centering
  \caption{
  }
  \includegraphics[width=\textwidth]{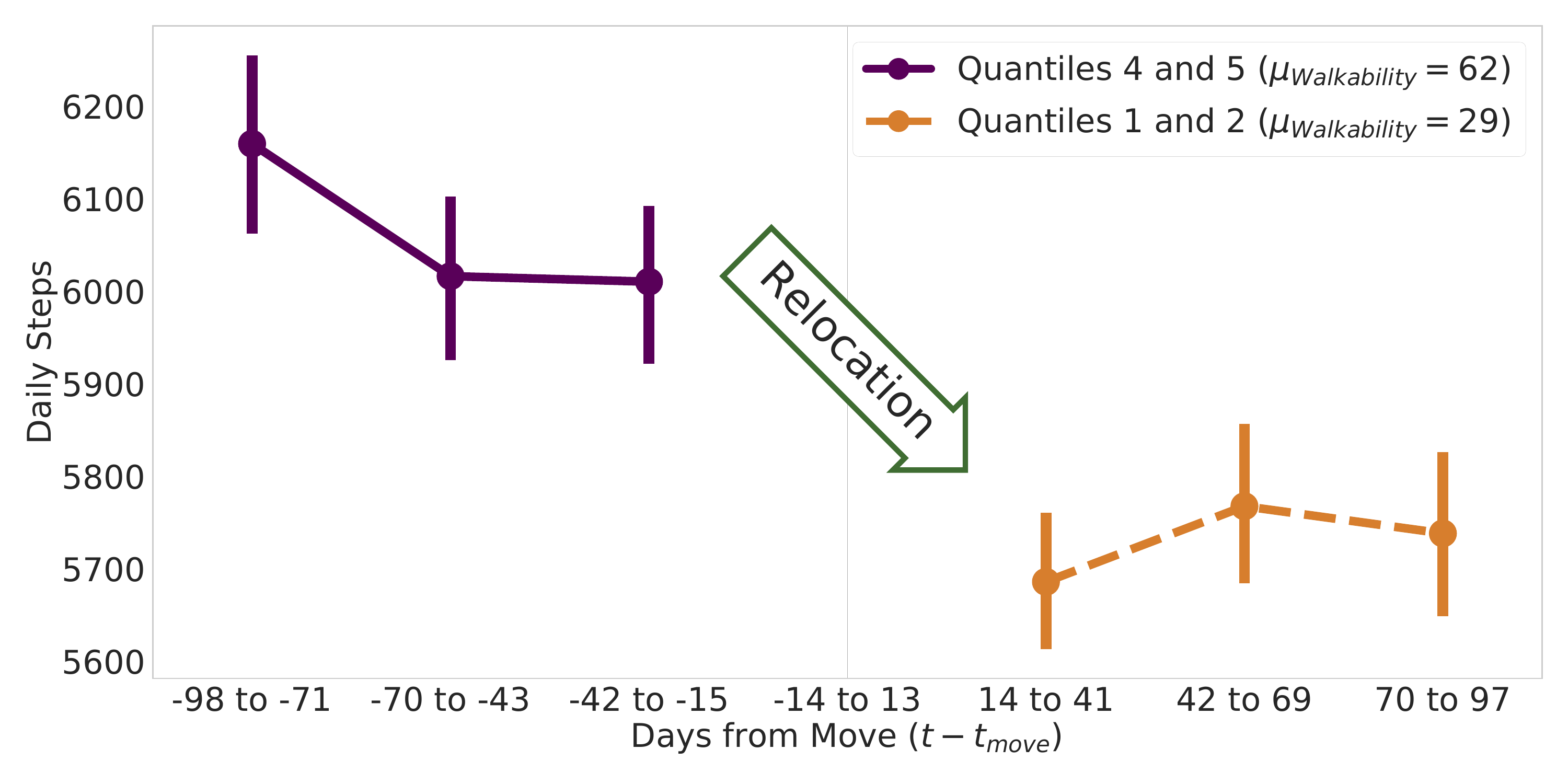}
  \label{fig:si_43_10_90d_monthavg}
\end{subfigure}
\vspace{-5mm}
\caption{
  \textbf{Relocations are not uniformly distributed across the year and activity levels need to be adjusted to exclude seasonal effects (Methods).}
  \textbf{a}, Histogram of number of days with tracked physical activity is non-uniform with more relocations in December and January.
  \textbf{b}, Changes in physical activity when relocating from highly walkable to less walkable built environments, without adjusting for the number of relocations across seasons.
  We observe clearly changing physical activity levels before and after relocation that may be explained by the more moderate weather in fall and spring versus winter.
  \textbf{c},  After adjusting for seasonal effects by weighting all months equally, physical activity levels before and after relocation are more stable.
} 
\label{fig:why_month_average}

\end{figure}

\begin{figure}[tb]
\vspace{-5mm}
\begin{subfigure}{.48\textwidth}
  \centering
  \caption{Moving from San Diego, CA.}
  \includegraphics[width=\textwidth]{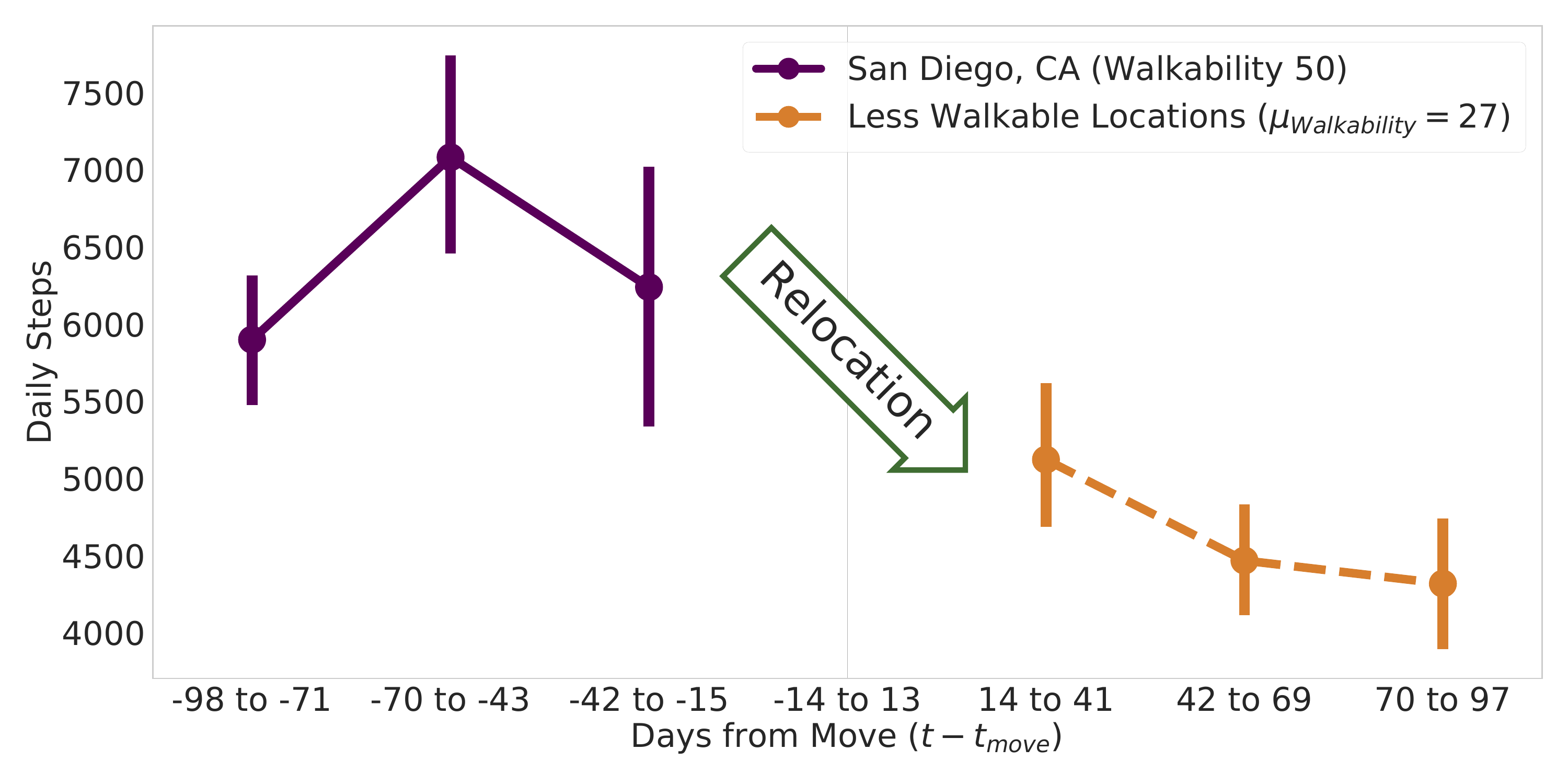}
     \label{fig:si_90d_from_SAN}
\end{subfigure}\hfill
\begin{subfigure}{.48\textwidth}
  \centering
  \caption{Moving to San Diego, CA.}
  \includegraphics[width=\textwidth]{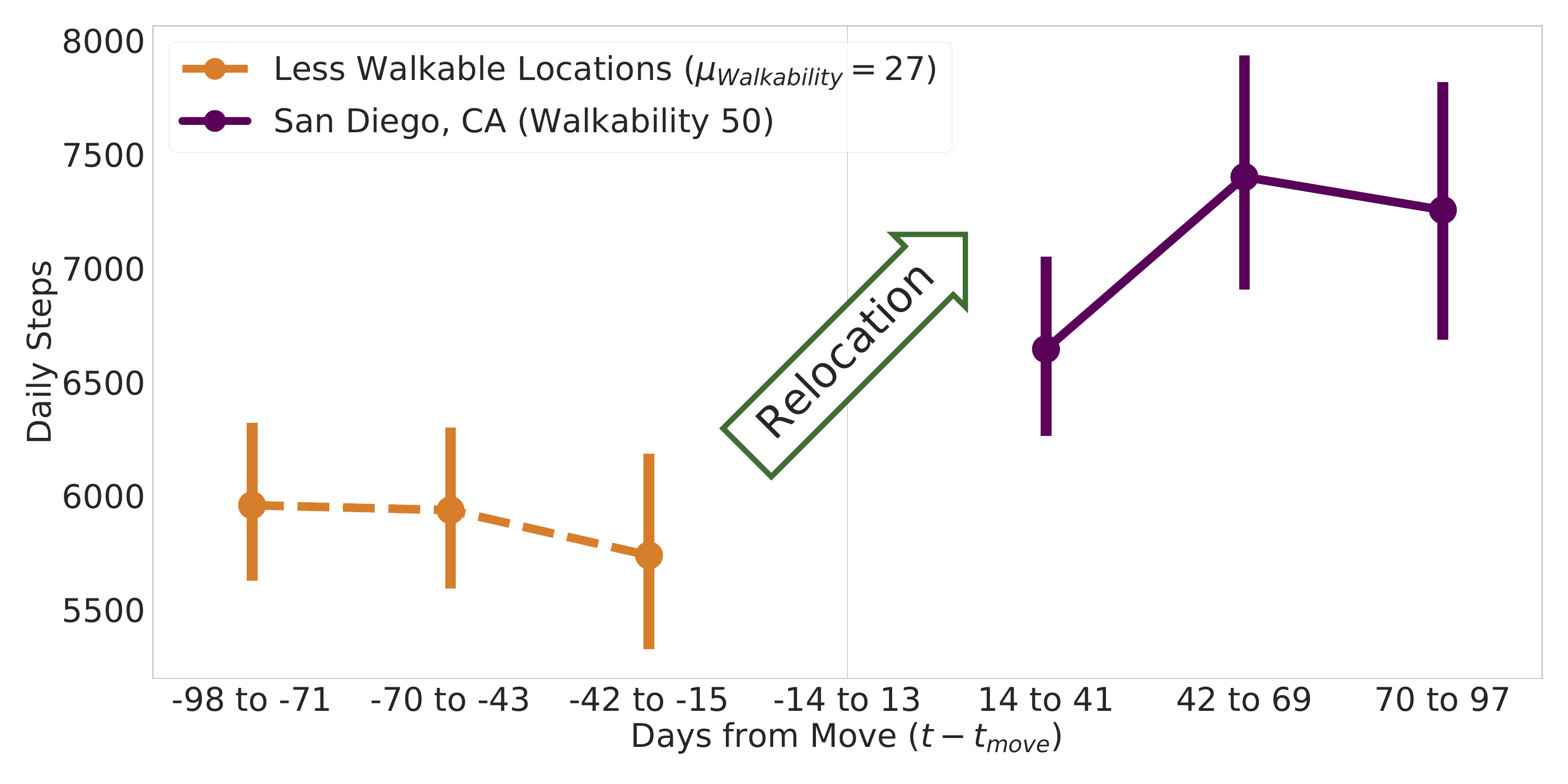}
     \label{fig:si_90d_to_SAN}
\end{subfigure}\\
\begin{subfigure}{.48\textwidth}
  \centering
  \caption{Moving from San Francisco, CA.}
  \includegraphics[width=\textwidth]{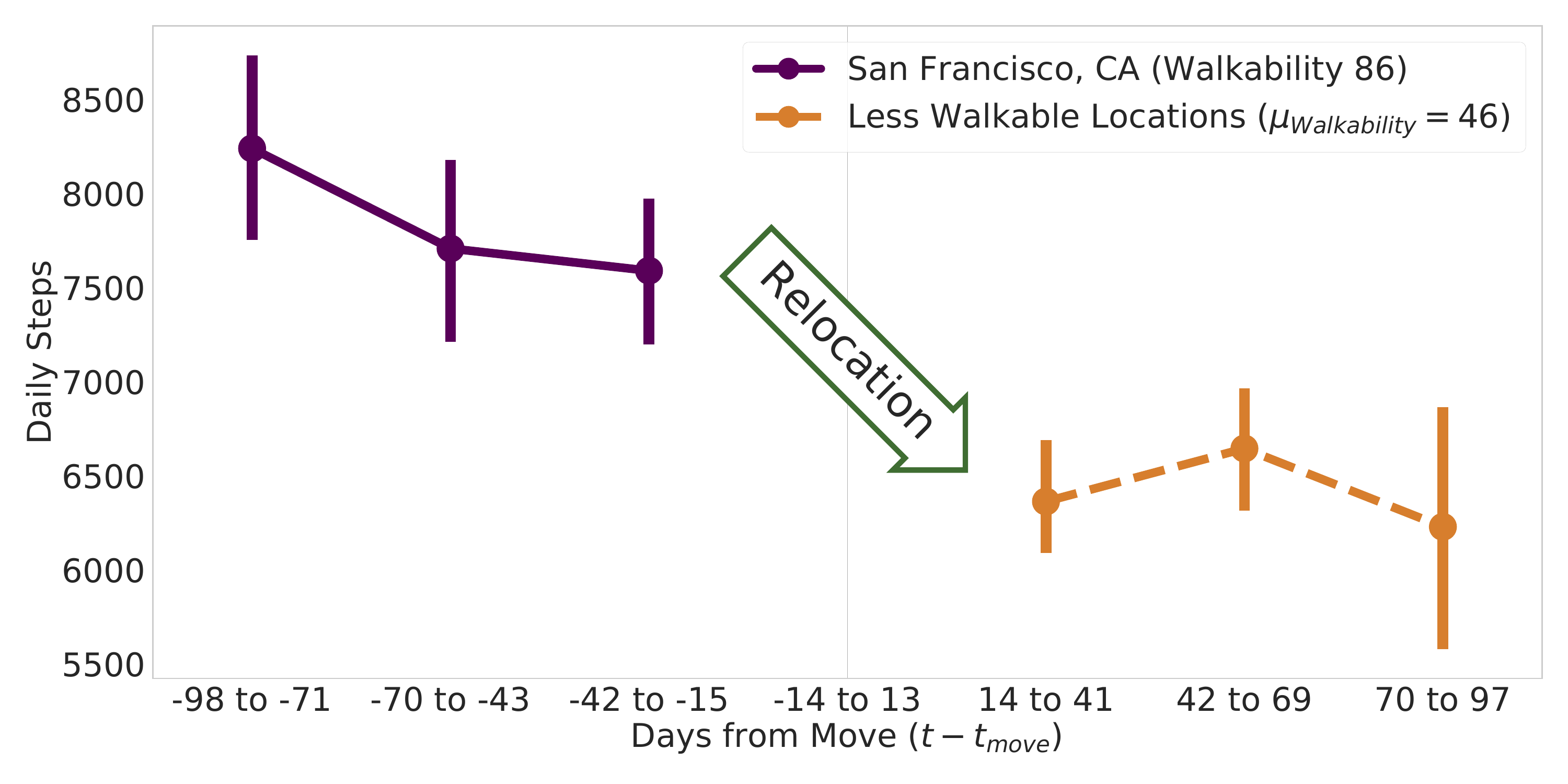}
     \label{fig:si_90d_from_TUS}
\end{subfigure}\hfill
\begin{subfigure}{.48\textwidth}
  \centering
  \caption{Moving to San Francisco, CA.}
  \includegraphics[width=\textwidth]{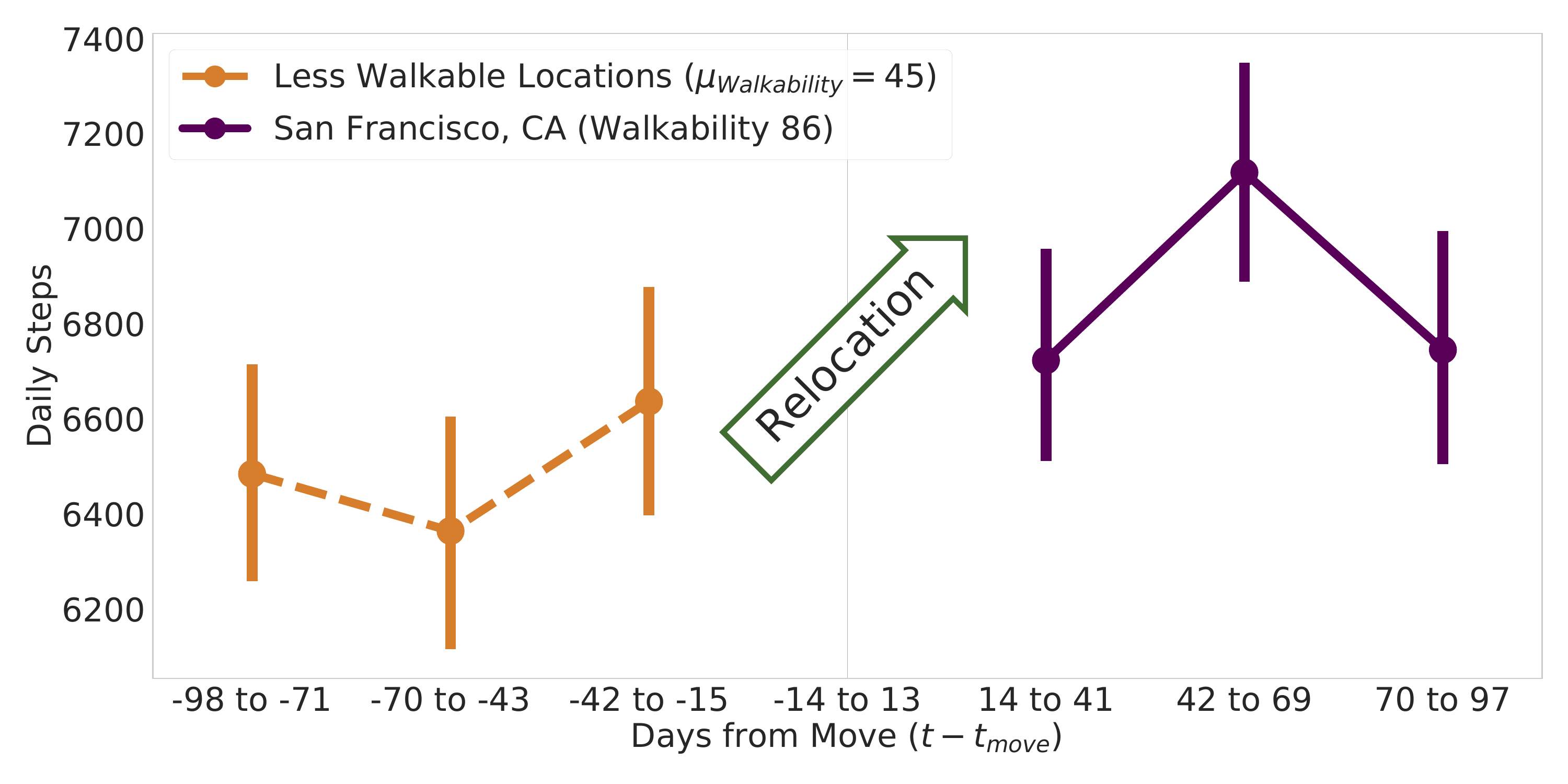}
     \label{fig:si_90d_to_TUS}
\end{subfigure}\\
\begin{subfigure}{.48\textwidth}
  \centering
  \caption{Moving from Baltimore, MD.}
  \includegraphics[width=\textwidth]{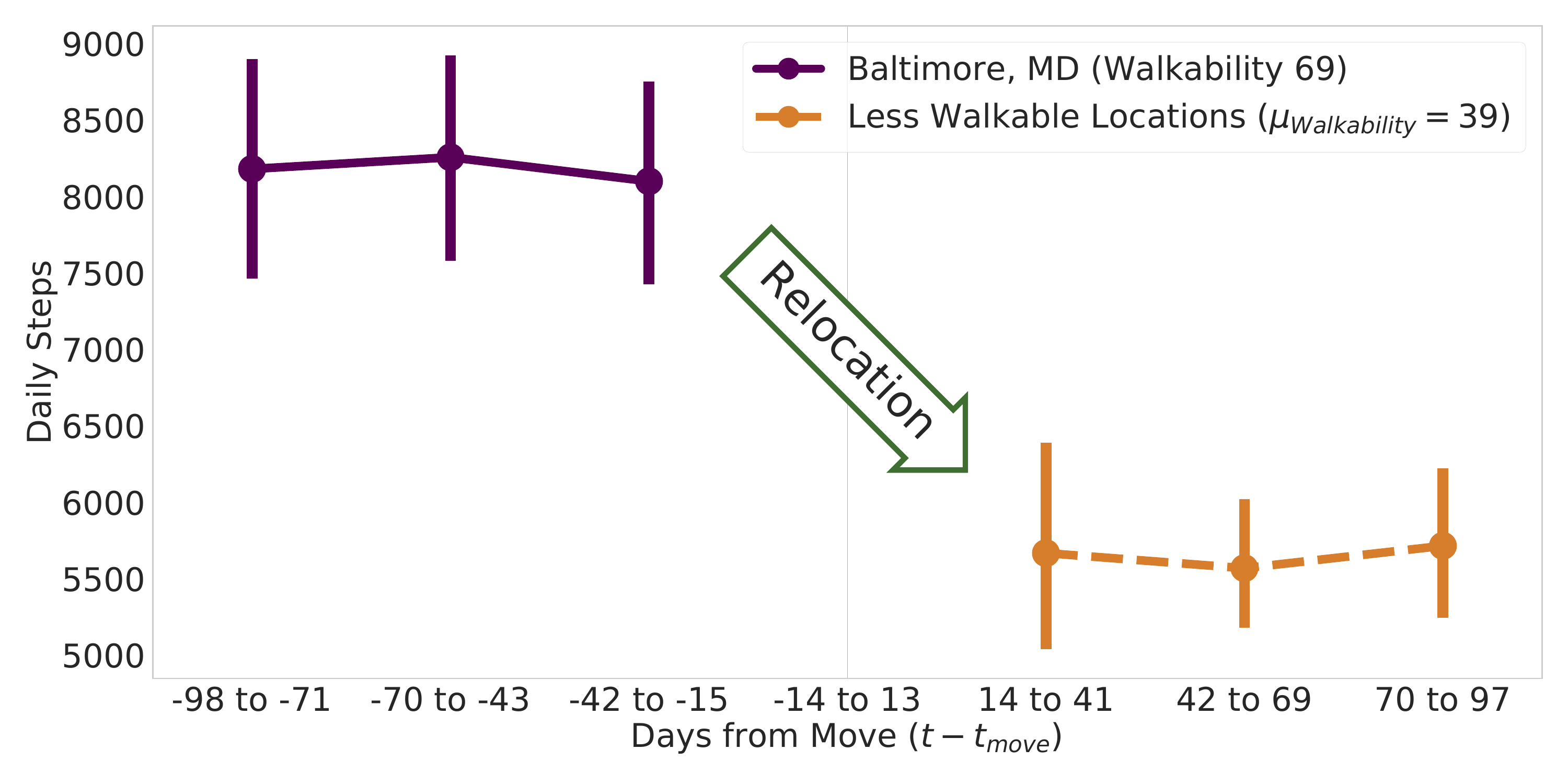}
     \label{fig:si_90d_from_BWI}
\end{subfigure}\hfill
\begin{subfigure}{.48\textwidth}
  \centering
  \caption{Moving to Baltimore, MD.}
  \includegraphics[width=\textwidth]{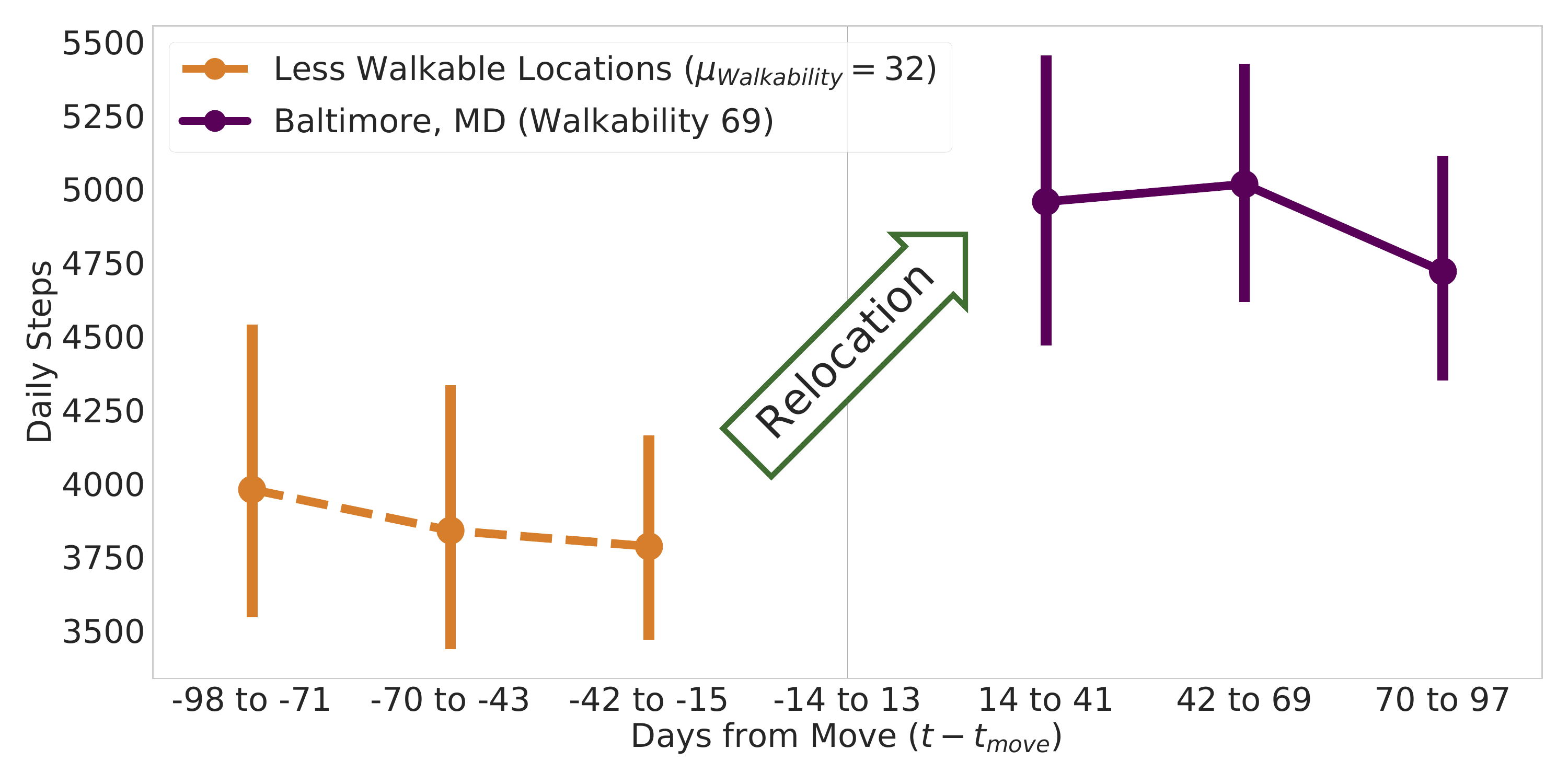}
     \label{fig:si_90d_to_BWI}
\end{subfigure}
\vspace{-5mm}
\caption{
\textbf{Changes in physical activity levels following relocation are still observed after three months.} 
Examples show physical activity levels for \participants~moving from/to San Diego, CA, San Francisco, CA, and Baltimore, MD.
The observation period is extended from 30 days to 90 days pre- and post-relocation. 
Observed changes in physical activity levels suggest that built environment influences persist over at least three months and may lead to sustained long-term behavior change.
} 
\label{fig:180_day_effects}
\end{figure}

\begin{figure}[t]
    \vspace{-30pt}
    \centering
    {\scriptsize \spacing{1.25}\fontsize{7}{8}\selectfont
        
        \iffigures
        \includegraphics[width=.95\textwidth]{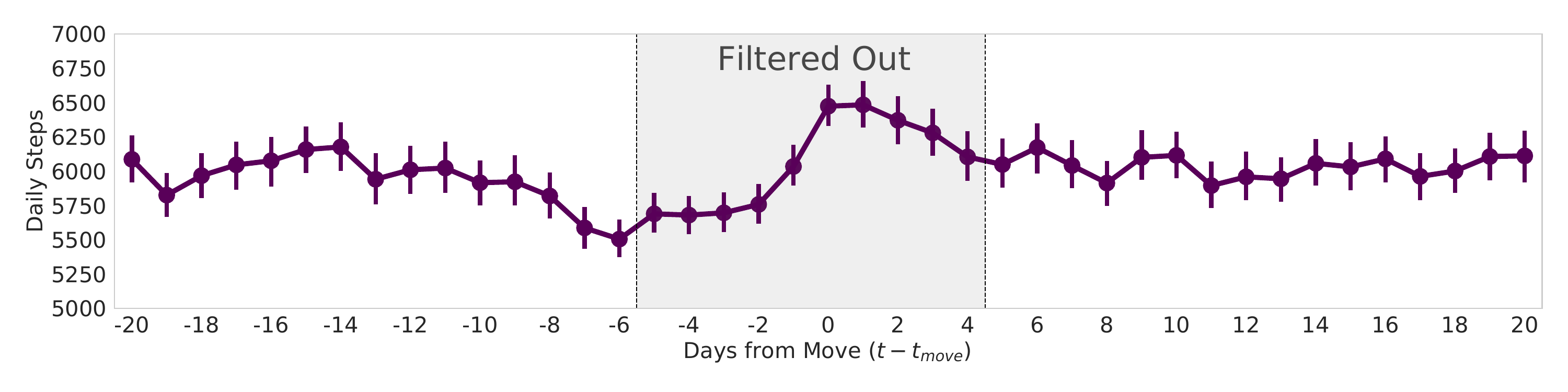} %
        \fi
        
        \caption{
        \textbf{Five days immediately before and after relocation are filtered out to exclude effects from the relocation process itself.}
        Outside this interval, \participants' physical activity levels were relatively stable. 
        Figure depicts relocations to similarly walkable locations (walkability score difference between -16 and 16).
        }
        \label{fig:si_why_remove_mid}
    } %
\end{figure}

\begin{figure}[tb]
\vspace{-5mm}
\begin{subfigure}{\textwidth}
  \centering
  \caption{}
  \includegraphics[width=.35\textwidth]{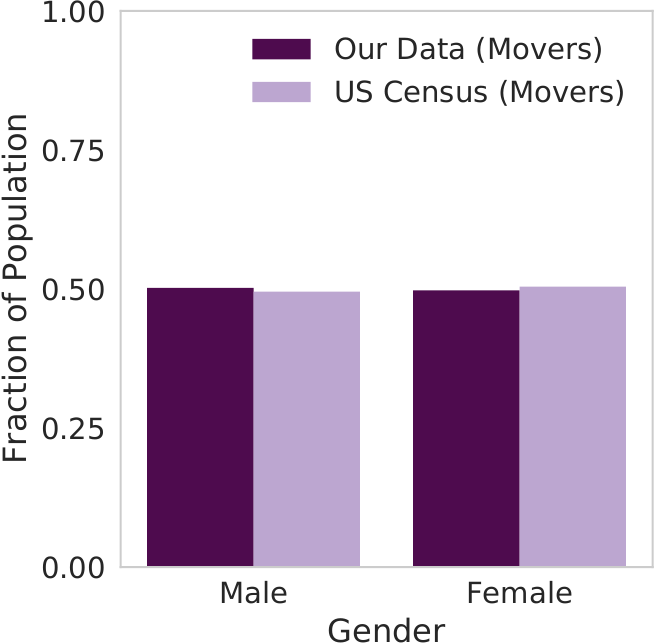}
     \label{fig:si_us_census_male_female_comparison}
\end{subfigure}\\
\begin{subfigure}{\textwidth}
  \centering
  \caption{}
  \includegraphics[width=.70\textwidth]{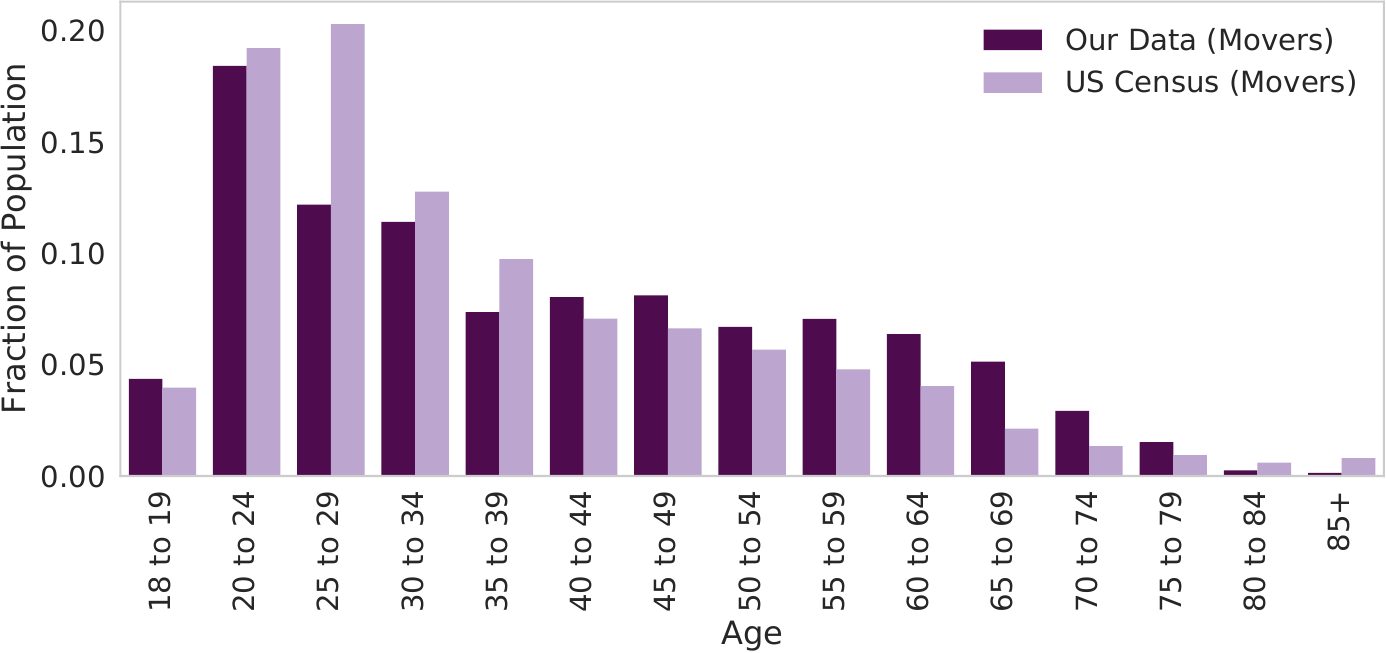}
     \label{fig:si_us_census_comparison}
\end{subfigure}

\caption{
\textbf{Demographics of relocating \participants~(Movers) in our dataset compared to U.S. Census estimates for movers.}
While there are no significant differences in the gender distribution (49.8\% female vs. 50.4\% female, $P = 0.423$, Z-test), 
we find a slightly lower age in our data for movers compared to the US Census reported movers (36.0 vs 37.7 median age), that we corrected for in the simulation experiment (Figure~\ref{fig:panel3}f, Figure~\ref{fig:si_new_simulation_results}). 
}
\label{fig:si_azumio_vs_census}
\end{figure}

\begin{figure}[tb]
\vspace{-5mm}
\begin{subfigure}{0.48\textwidth}
  \centering
  \caption{}
  \includegraphics[width=\textwidth]{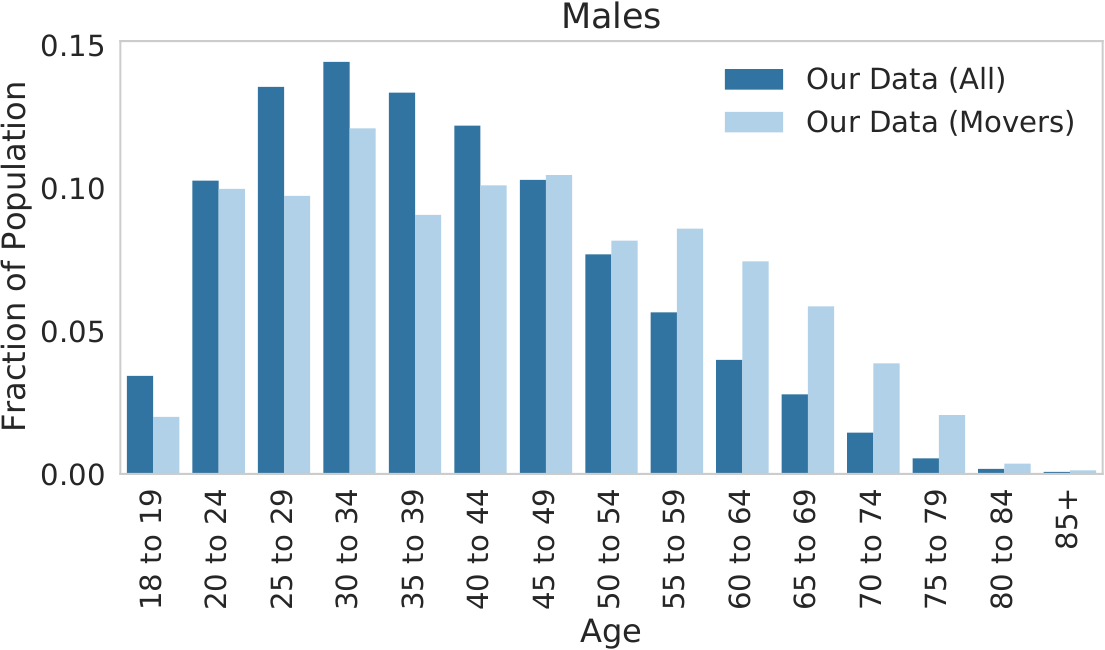}
     \label{fig:si_all_azumio_vs_movers_male_age_hist}
\end{subfigure}\hfill
\begin{subfigure}{0.48\textwidth}
  \centering
  \caption{}
  \includegraphics[width=\textwidth]{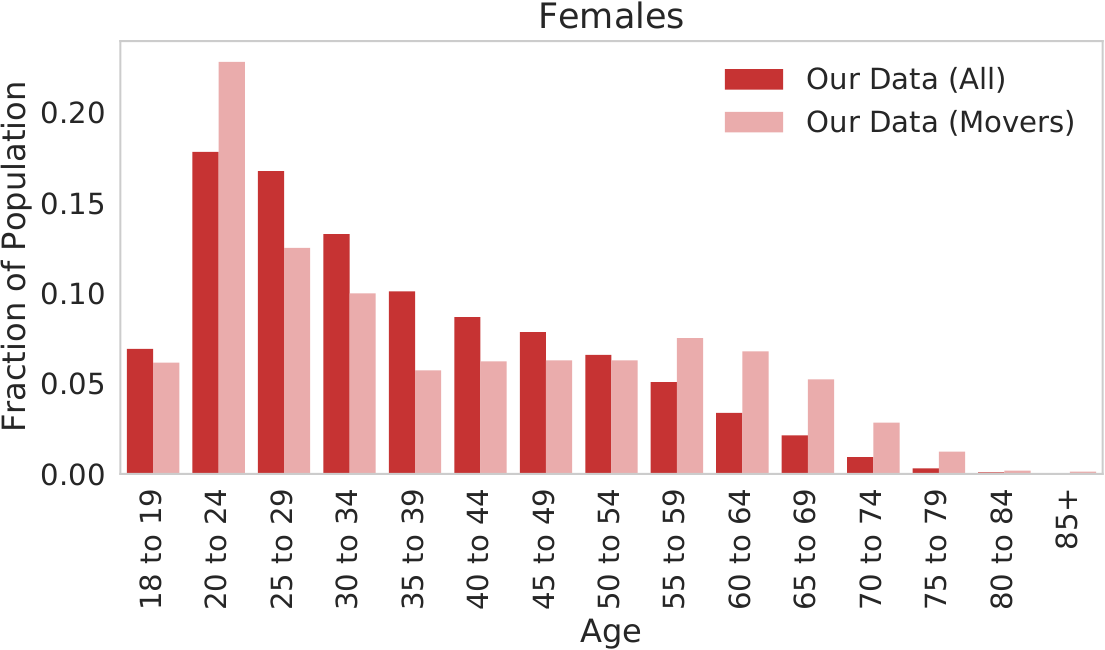}
     \label{fig:si_all_azumio_vs_movers_female_age_hist}
\end{subfigure}\\
\begin{subfigure}{0.48\textwidth}
  \centering
  \caption{}
  \includegraphics[width=\textwidth]{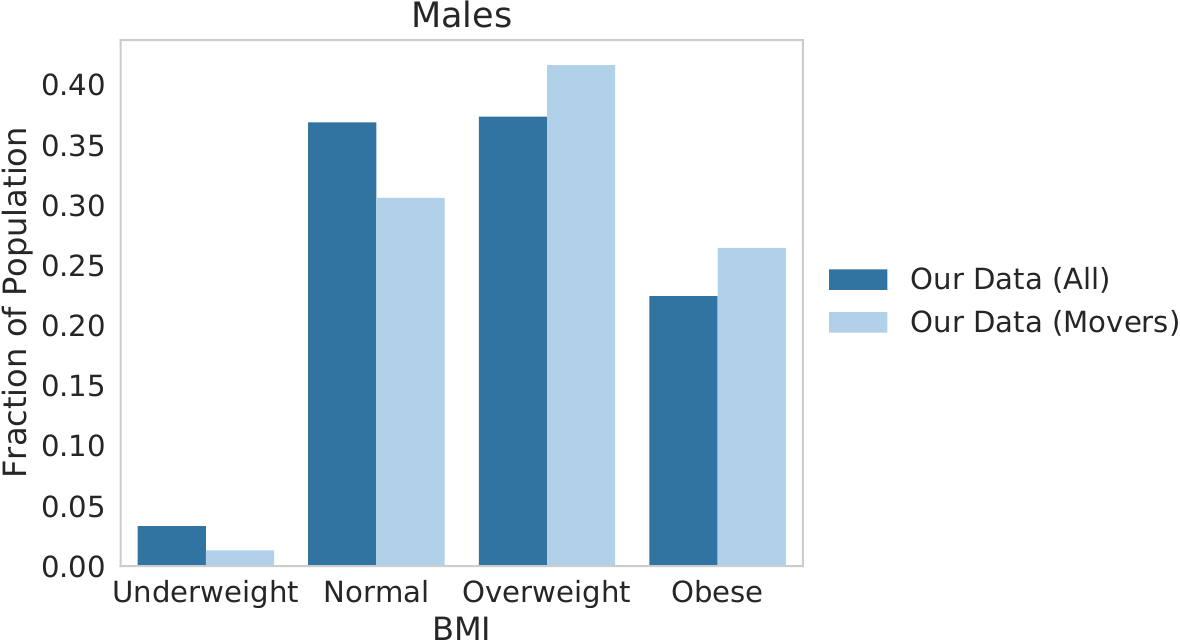}
     \label{fig:si_all_azumio_vs_movers_male_bmi_hist}
\end{subfigure}\hfill
\begin{subfigure}{0.48\textwidth}
  \centering
  \caption{}
  \includegraphics[width=\textwidth]{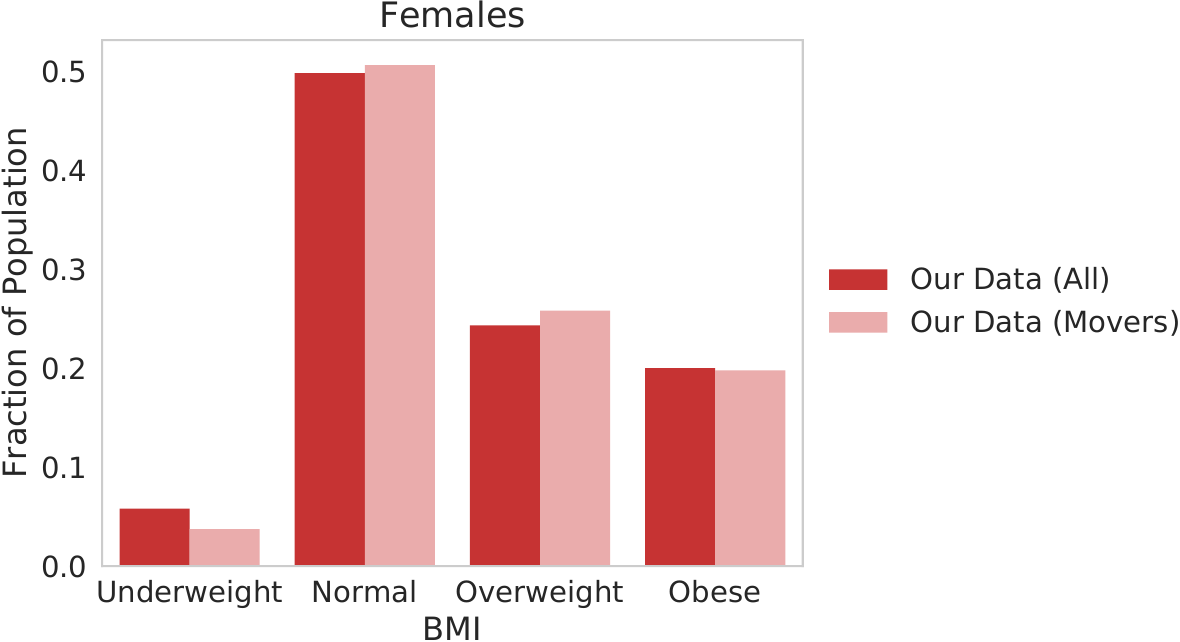}
     \label{fig:si_all_azumio_vs_movers_female_bmi_hist}
\end{subfigure}\\
\begin{subfigure}{0.48\textwidth}
  \centering
  \caption{}
  \includegraphics[width=\textwidth]{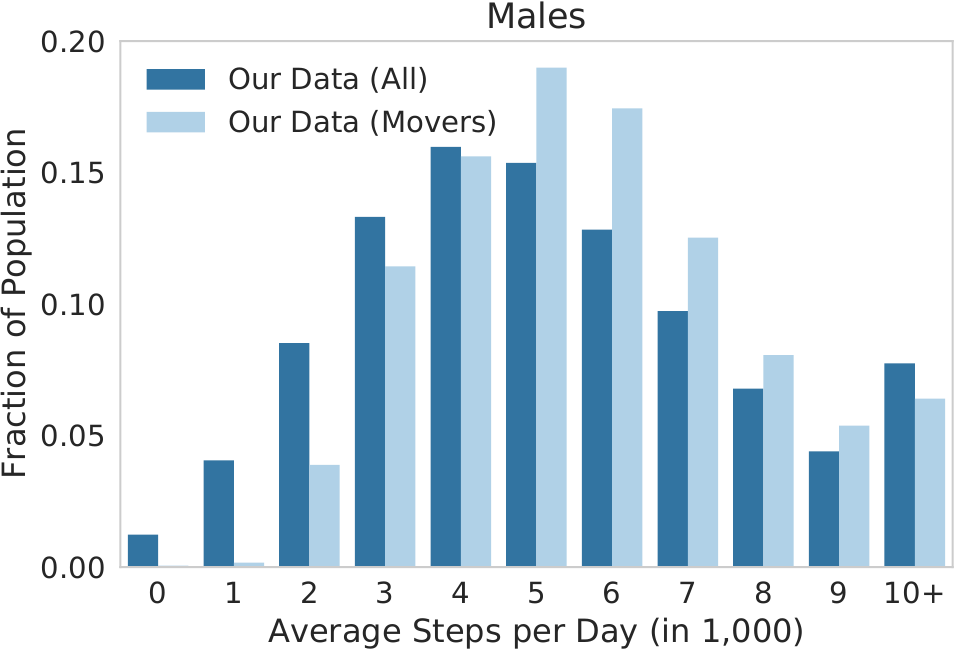}
     \label{fig:si_all_azumio_vs_movers_male_PA_hist}
\end{subfigure}\hfill
\begin{subfigure}{0.48\textwidth}
  \centering
  \caption{}
  \includegraphics[width=\textwidth]{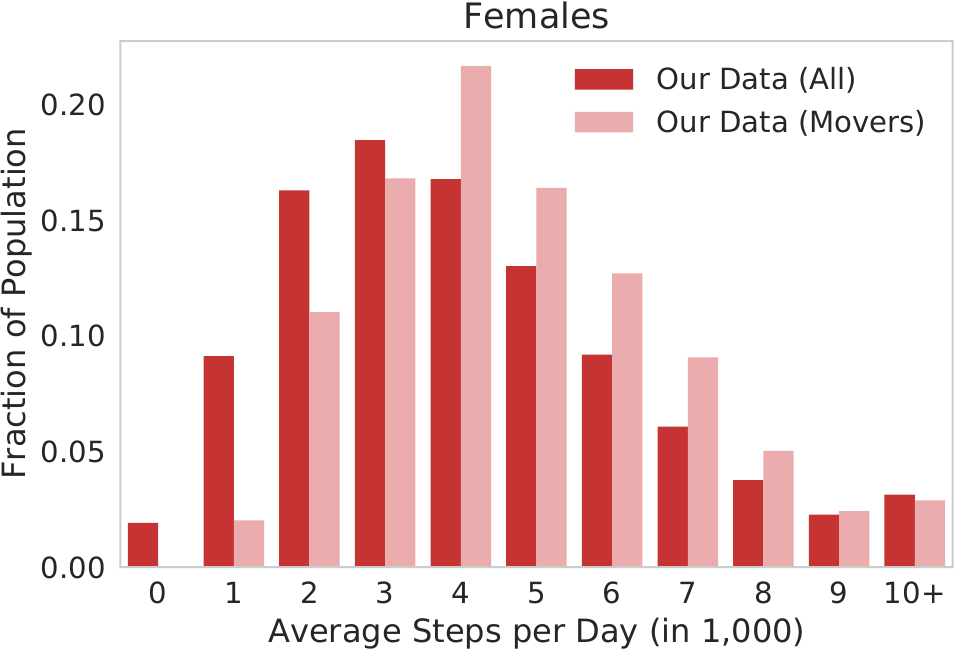}
     \label{fig:si_all_azumio_vs_movers_female_PA_hist}
\end{subfigure}

\caption{
\textbf{Comparison of relocating \participants~(movers) to all \participants~in our dataset to understand potential selection effects.}
Within the study population, we find that movers and non-movers (\ie, relocating and non-relocating \participants) tend to be
\textbf{a-b,} close in age (43.8 vs 37.9 and 38.5 vs 33.7 average age for men and women, respectively), 
\textbf{c-d,} and weight status (68.1 vs 59.8 and 45.6 vs 44.3 percent overweight and obese for men and women, respectively). 
\textbf{e-f,} However, movers were generally more physically active than non-movers (6,284 vs 5,825 and 5,279 vs 4,635 average daily steps for men and women, respectively). 
}
\label{fig:si_movers_vs_nonmovers}
\end{figure}

\begin{figure}[tb]
\vspace{-5mm}
\begin{subfigure}{0.48\textwidth}
  \centering
  \caption{}
  \includegraphics[width=\textwidth]{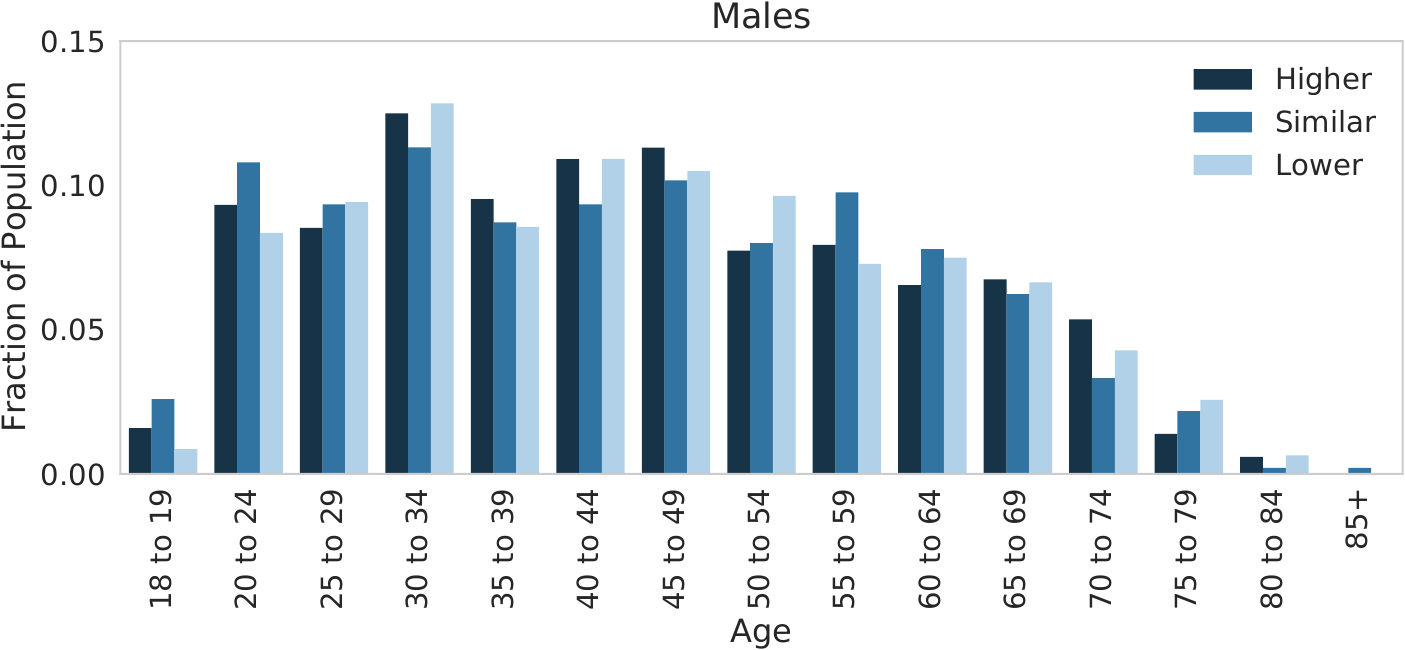}
     \label{fig:si_movers_vs_ws_diff_male_age_hist}
\end{subfigure}\hfill
\begin{subfigure}{0.48\textwidth}
  \centering
  \caption{}
  \includegraphics[width=\textwidth]{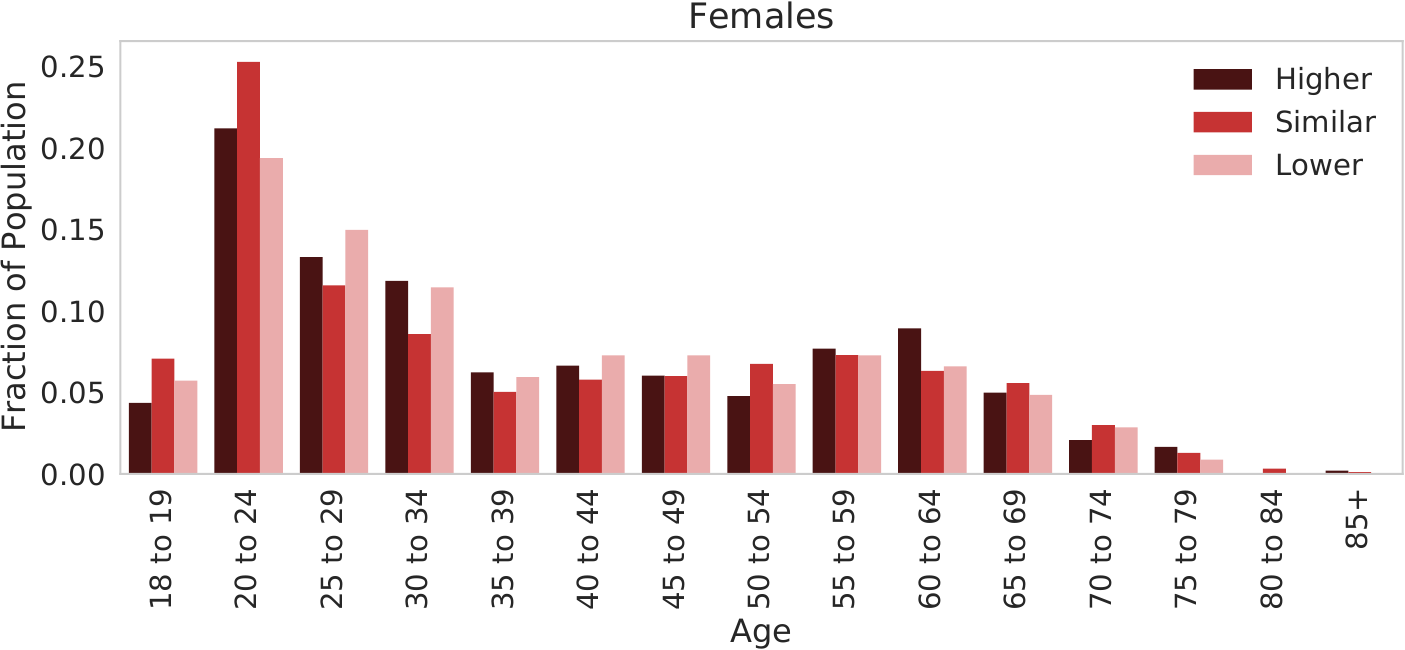}
     \label{fig:si_movers_vs_ws_diff_female_age_hist}
\end{subfigure}\\
\begin{subfigure}{0.48\textwidth}
  \centering
  \caption{}
  \includegraphics[width=\textwidth]{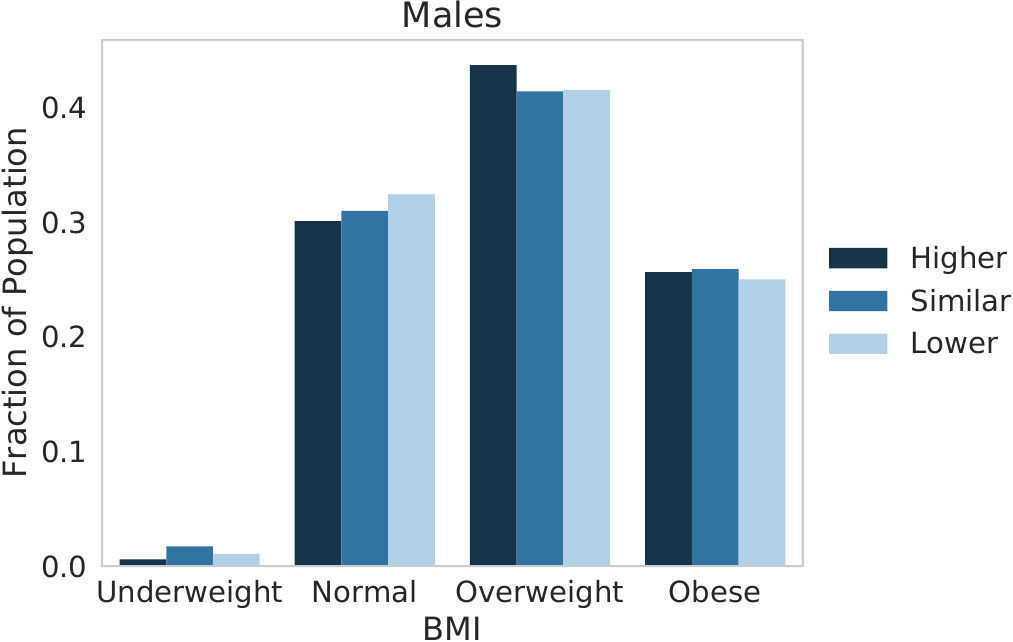}
     \label{fig:si_movers_vs_ws_diff_male_bmi_hist}
\end{subfigure}\hfill
\begin{subfigure}{0.48\textwidth}
  \centering
  \caption{}
  \includegraphics[width=\textwidth]{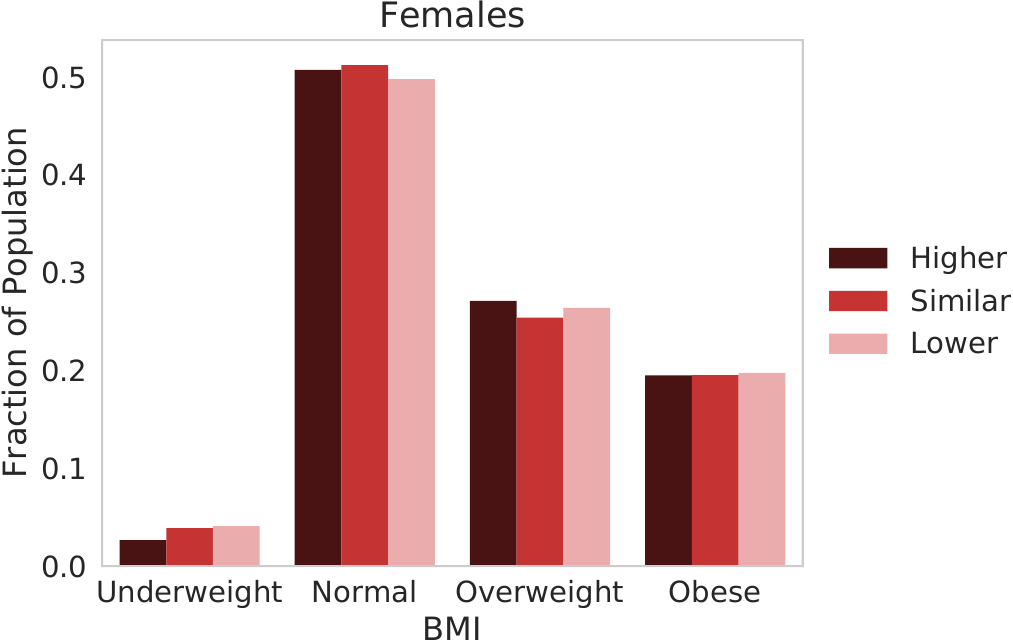}
     \label{fig:si_movers_vs_ws_diff_female_bmi_hist}
\end{subfigure}\\
\begin{subfigure}{0.48\textwidth}
  \centering
  \caption{}
  \includegraphics[width=\textwidth]{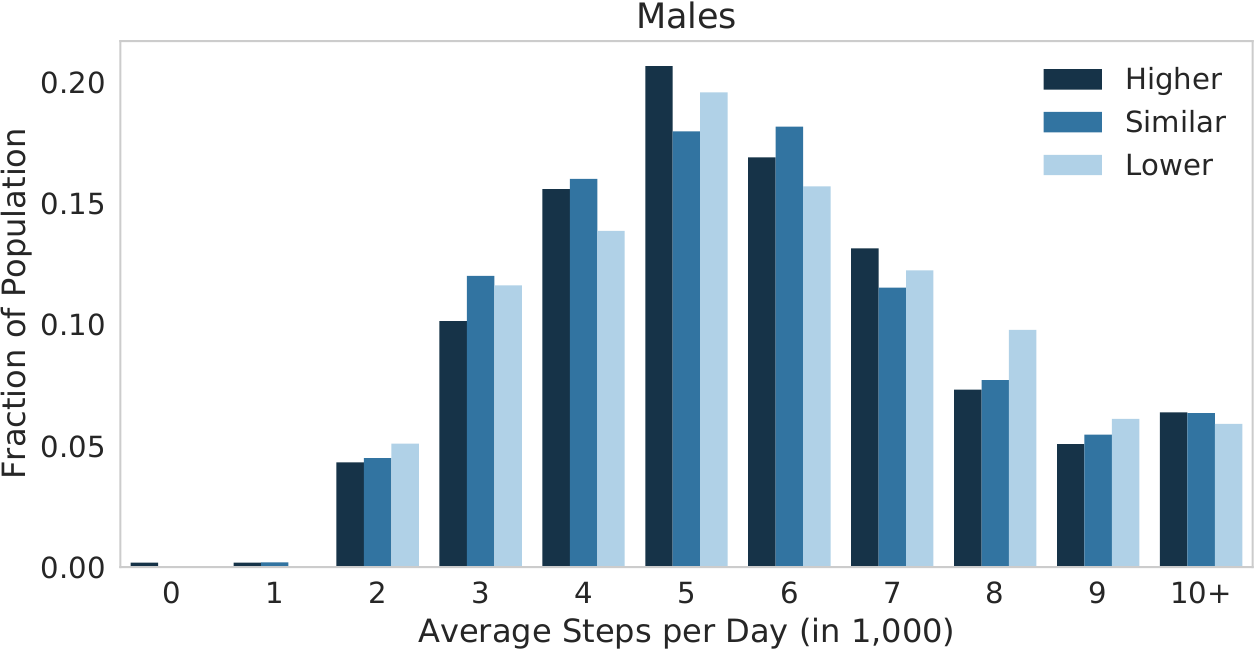}
     \label{fig:si_movers_vs_ws_diff_male_PA_hist}
\end{subfigure}\hfill
\begin{subfigure}{0.48\textwidth}
  \centering
  \caption{}
  \includegraphics[width=\textwidth]{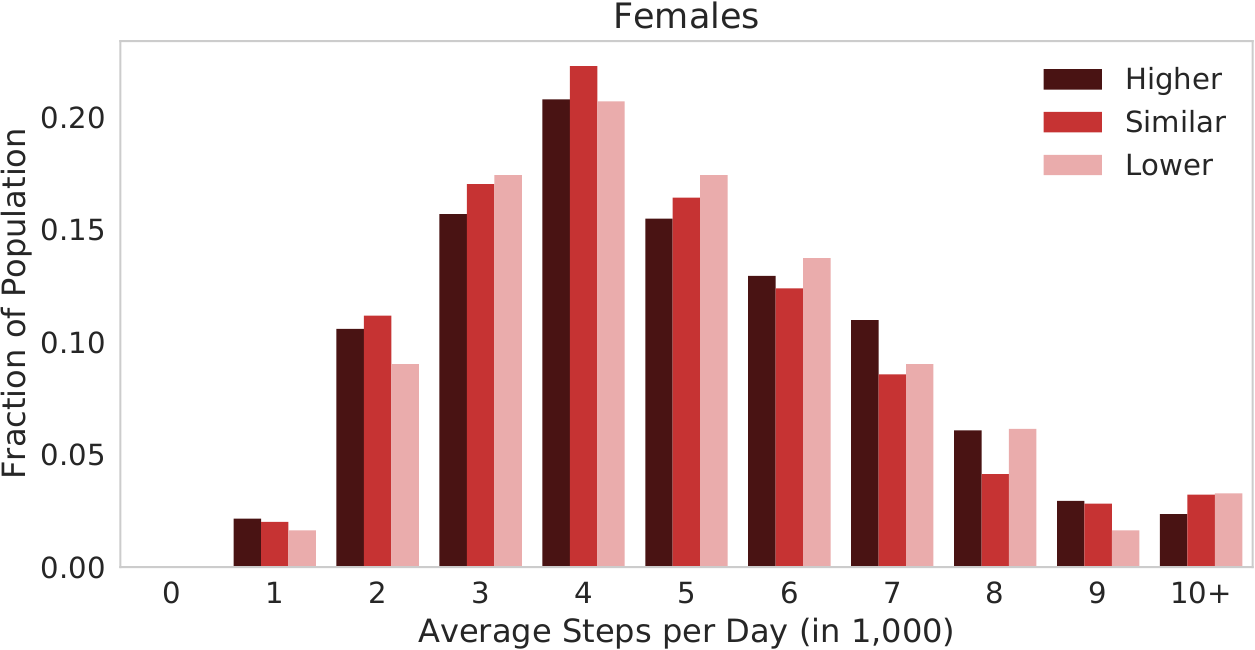}
     \label{fig:si_movers_vs_ws_diff_female_PA_hist}
\end{subfigure}

\caption{
\textbf{Comparison of different groups of relocating \participants~(movers) to understand potential selection effects.}
Within the study population, we find that movers to higher (greater than 16 walkability points), similar (within 16 walkability points), and lower (less than -16 walkability points) walkability locations tend to be
\textbf{a-b,} close in age (44.4 vs 43.7 vs 45.0 and 39.0 vs 38.3 vs 38.3 average age for men and women moving to higher, similar, and lower walkability locations, respectively), 
\textbf{c-d,} weight status (69.3 vs 67.3 vs 66.5 and 46.6 vs 44.9 vs 46.1 percent overweight and obese for men and women moving to higher, similar, and lower walkability locations, respectively),
\textbf{e-f,} and baseline physical activity levels (6,257 vs 6,229 vs 6,301 and 5,422 vs 5,239 vs 5,405 average daily steps for men and women moving to higher, similar, and lower walkability locations, respectively).
}
\label{fig:si_movers_vs_ws_diff}
\end{figure}

\begin{figure}[tb]
\begin{subfigure}{\textwidth}
  \centering
  \includegraphics[width=0.85\textwidth]{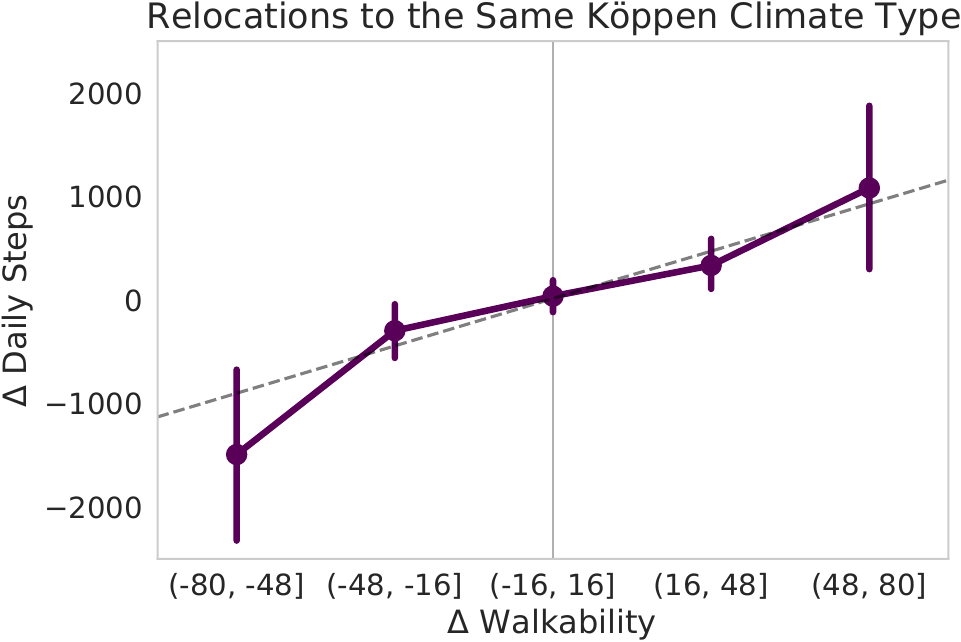}
  \label{fig:steps_vs_binned_ws_same_koppen}
\end{subfigure}
\vspace{-5mm}
\caption{
\textbf{Relationship between city walkability and physical activity holds for relocations to the same K\"{o}ppen climate type.}
We find that that relocations to more walkable cities are associated with significant increases in physical activity across moves to the same K\"{o}ppen climate type. 
These results suggest that our main result---city walkability impacts physical activity---is independent of any potential climate bias in our sample (i.e., moves to a more favorable climate for physical activity are not driving the differences we observe).
}
\label{fig:si_main_effect_by_koppen}
\end{figure}

\begin{figure}[tb]
\begin{subfigure}{\textwidth}
  \centering
  \includegraphics[width=0.85\textwidth]{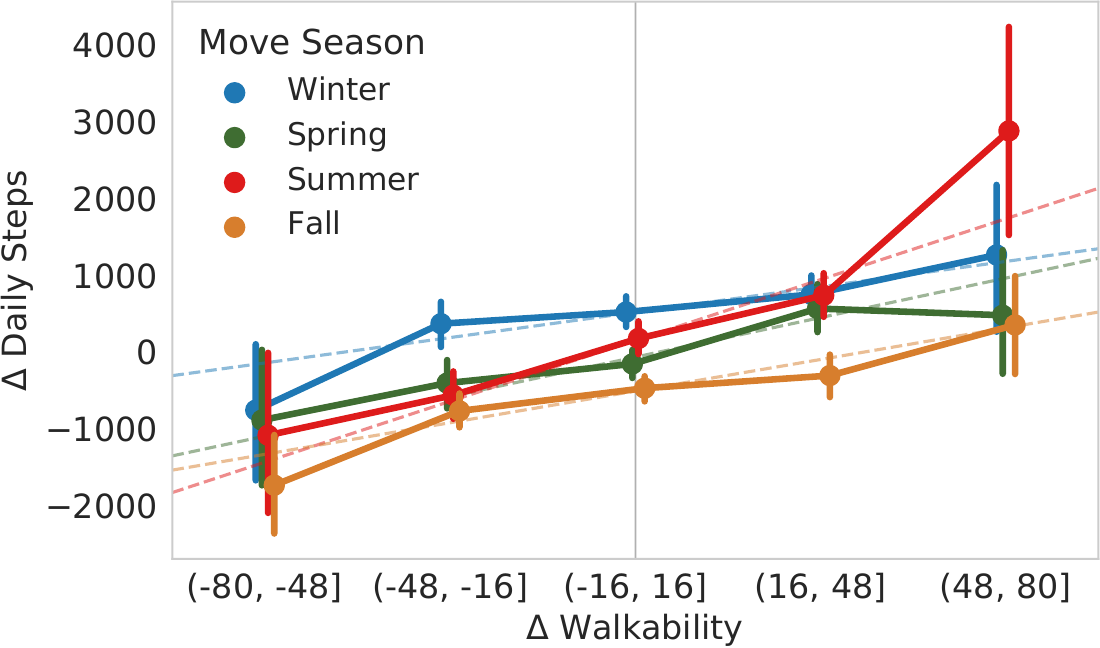}
  \label{fig:steps_vs_binned_ws_season}
\end{subfigure}
\vspace{-5mm}
\caption{
\textbf{Relationship between city walkability and physical activity holds for relocations within any given season.}
We find that that relocations to more walkable cities are associated with significant increases in physical activity across moves that occur within any given season (all $P<10^{-3}$). 
These results suggest that our main result---city walkability impacts physical activity---is independent of any potential temporal bias in our sample (i.e., moves during a time that is more favorable for physical activity are not driving the differences we observe).
}
\label{fig:si_main_effect_by_season}
\end{figure}

\begin{figure}[tb]
\begin{subfigure}{\textwidth}
  \centering
  \includegraphics[width=\textwidth]{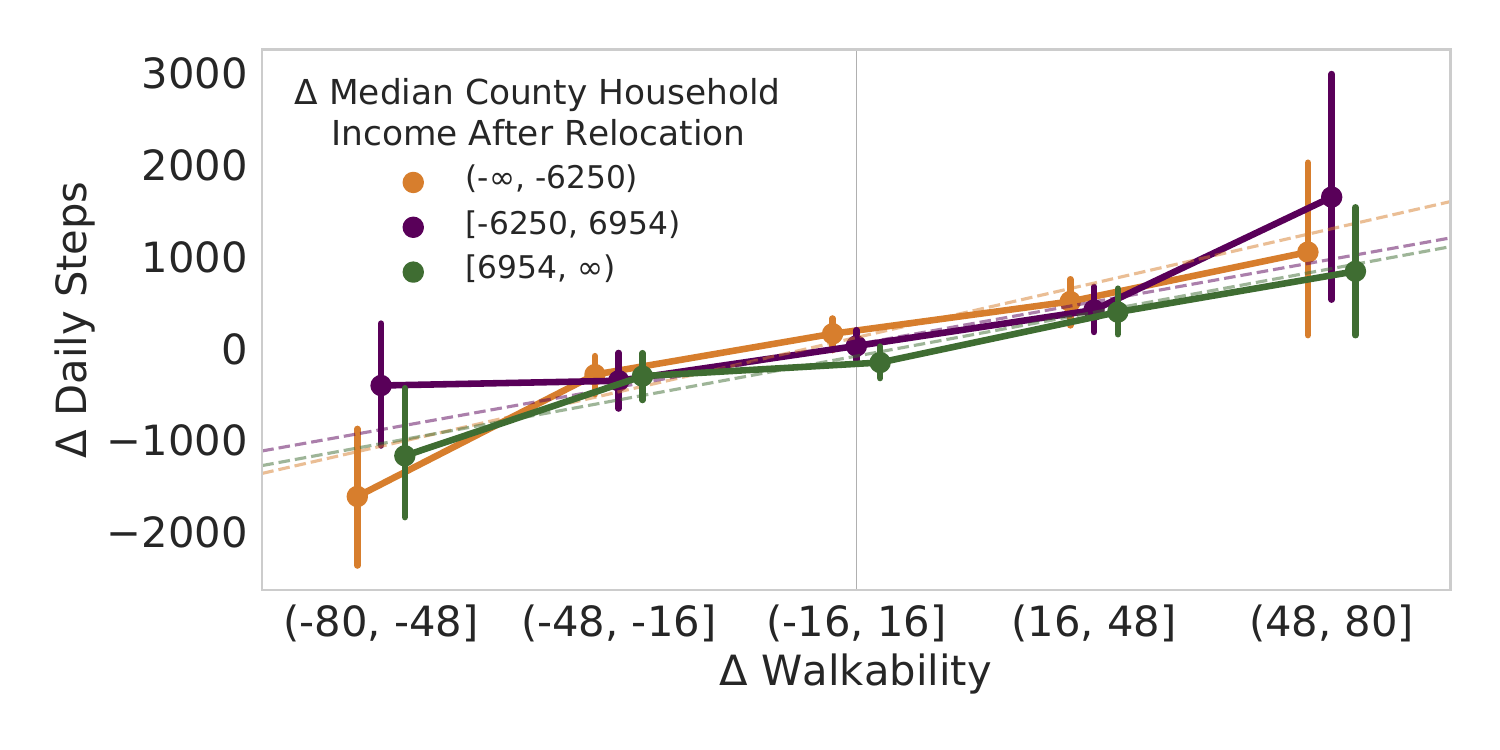}
  \label{fig:steps_vs_binned_ws_strat_income}
\end{subfigure}
\vspace{-20mm}
\caption{
\textbf{Relationship between city walkability and physical activity holds within U.S. cities of similar income.}
We find that that relocations to more walkable cities are associated with significant increases in physical activity across all three groups (increasing, similar, and decreasing median county household income in USD). 
These results suggest that our main result---city walkability impacts physical activity---is robust to potential socioeconomic bias in our sample.
}
\label{fig:si_main_effect_by_income}
\end{figure}

\begin{figure}[tb]
\begin{subfigure}{\textwidth}
  \centering
  \includegraphics[width=0.80\textwidth]{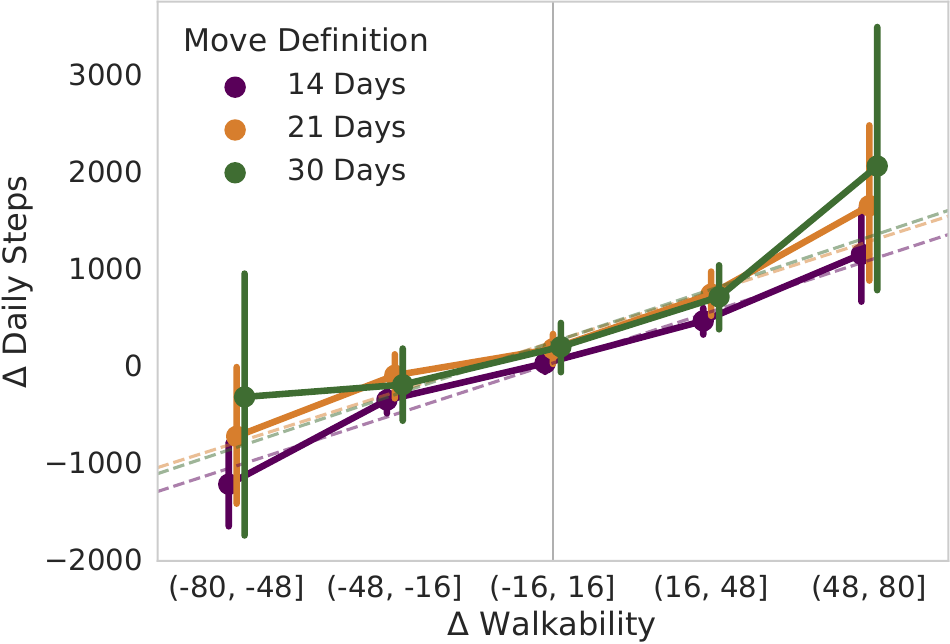}
  \label{fig:stratified_by_move_def}
\end{subfigure}
\vspace{-10mm}
\caption{
\textbf{Different definitions of relocation lead to highly consistent results.}
After relocation, \participants~are required to stay in the new location for at least 14, 21, or 30 days. 
We find that all of these definitions lead to highly consistent results as most relocating \participants~stay for substantially longer periods of time (median 81 days).
In the rest of the paper, we use the 14 day definition (purple).
}
\label{fig:fig2a_by_move_def}
\end{figure}

\begin{figure}[tb]
\begin{subfigure}{0.48\textwidth}
  \centering
  \includegraphics[width=\textwidth]{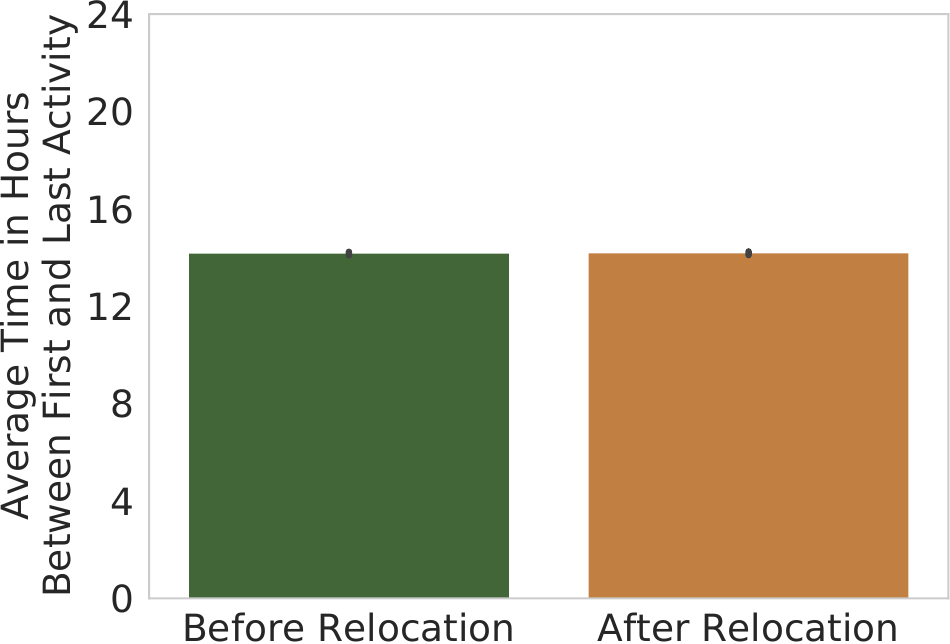}
  \caption{}
  \label{fig:active_daily_hours}
\end{subfigure}
\hfill
\begin{subfigure}{0.49\textwidth}
  \centering
  \includegraphics[width=\textwidth]{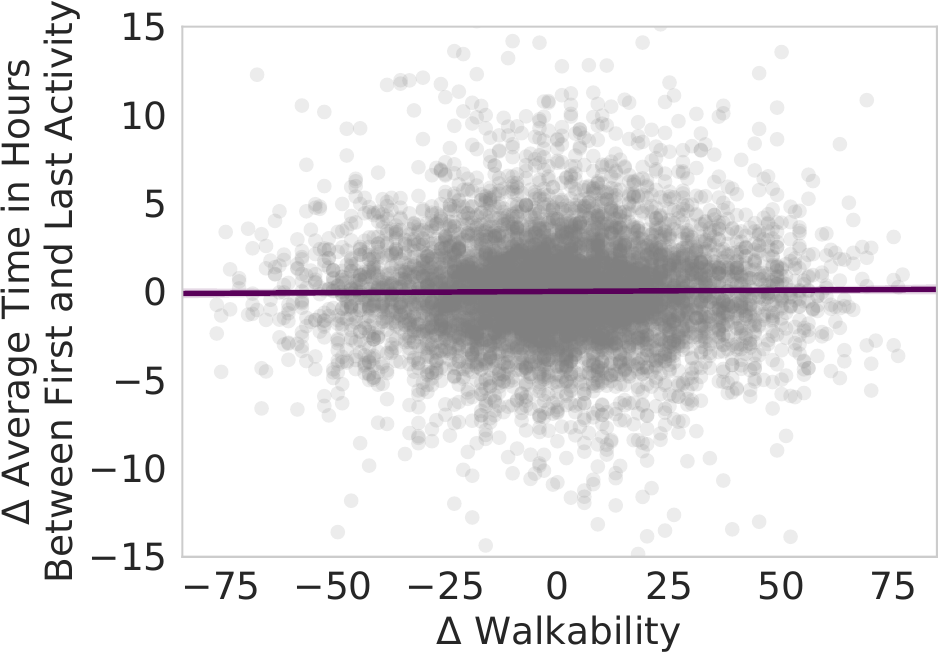}
  \caption{}
  \label{fig:hours_used_diff}
\end{subfigure}
\vspace{-3mm}
\caption{
\textbf{Higher physical activity in more walkability cities is not explained by differences in estimated wear time.}
\Participants~have an average span of 14.2h between the first and last recorded step, our proxy for daily wear time (Methods).
\textbf{a,} Wear time estimates before and after relocation are 14.16 hours and 14.18 hours, respectively, with no significant difference ($P = 0.807$).
\textbf{b,} 
We find no significant association between relocation-induced difference in walkability and wear time.
The line shows the best linear fit using data from all relocations.
Its slope is not significantly different from zero (slope $0.0014$; $P = 0.371$). 
These results suggest that differences in recorded steps after relocation are due to actual differences in physical activity behavior and are not explained by differences in wear time.
}
\label{fig:si_weartime}
\end{figure}

\clearpage

\begin{table}
\centering
\caption{
Summary of demographic statistics for the study \participants~(Methods). 
Study observation period ranged from March 2013 to February 2016. 
Percentages are in parentheses.
NA refers to missingness in data.
}
\medskip
\begin{tabular}{lccc}
\toprule
Quantity & All \participants & Movers & Non-movers\\
\midrule
Total \participants~& \numtotalusers~(100.0) & \numusers~(100.0) & 2,106,864 (100.0)\\
Median Age & 32 & 36 & 32\\
Moves per \participant & - & Min: 1.0, Max: 9.0, Avg: 1.37 & -\\
\# Female & 413,373 (48.1) & 1,732 (49.8) & 411,641 (48.1)\\
\# Male & 446,406 (51.9) & 1,748 (50.2) & 444,658 (51.9)\\
\# gender NA & 1,252,509 (59.3) & 1,944 (35.8) & 1,250,565 (59.4)\\
\# Overweight & 347,964 (30.4) & 1,549 (33.2) & 346,415 (30.4)\\
\# Obese & 241,842 (21.2) & 1,073 (23.0) & 240,769 (21.2)\\
\# BMI NA & 969,538 (45.9) & 760 (14.0) & 968,778 (46.0)\\
\bottomrule
\end{tabular}
\label{table:table_demographics}
\end{table}

\clearpage

\begin{table}
\centering
\caption{
Summary of physical activity statistics for the study \participants~(Methods). 
Statistics are averages unless otherwise noted with standard deviations in parentheses.
}
\medskip
\begin{tabular}{lccc}
\toprule
Quantity & Value\\
\midrule
\# Daily Steps\\
\hspace{1cm} Overall & 5574 (3055) \\
\hspace{1cm} Before Move & 5559 (3059) \\
\hspace{1cm} After Move & 5588 (3051) \\
\# Minutes MVPA\\
\hspace{1cm} Overall & 103 (104) \\
\hspace{1cm} Before Move & 102 (101) \\
\hspace{1cm} After Move & 104 (106) \\
\% Population Meeting PA Guidelines\\
\hspace{1cm} Overall & 23.9 \\
\hspace{1cm} Before Move & 23.4 \\
\hspace{1cm} After Move & 24.4 \\
\# Days Tracked per Relocation (30 days)\\
\hspace{1cm} Total & \numdaysThirty \\
\hspace{1cm} Min & 2 \\
\hspace{1cm} Max & 51 \\
\hspace{1cm} Mean & 33.3 (12.7) \\
\# Days Tracked per Relocation (90 days)\\
\hspace{1cm} Total & \numdaysNinety \\
\hspace{1cm} Min & 1 \\
\hspace{1cm} Max & 153 \\
\hspace{1cm} Mean & 80.0 (39.4) \\
\bottomrule
\end{tabular}
\label{table:table_pa_stats}
\end{table}

\clearpage

\begin{table}
\centering
\caption{
Location and walkability statistics for all locations included in our study with at least 70 moving \participants~(sorted alphabetically).
Percentages are in parentheses.
}
\medskip
\begin{tabular}{lcccc}
\toprule
City & Walkability & \# Moving \participants & \# Female & \# Overweight \\
\midrule
Atlanta, GA & 48 & 140 & 46 (53.5) & 37 (30.6) \\
Austin, TX & 39 & 123 & 38 (49.4) & 37 (35.6) \\
Boston, MA & 81 & 106 & 34 (53.1) & 27 (32.1) \\
Charlotte, NC & 26 & 71 & 20 (51.3) & 20 (37.0) \\
Chicago, IL & 78 & 222 & 60 (48.4) & 61 (32.8) \\
Dallas, TX & 45 & 121 & 27 (34.2) & 46 (43.4) \\
Denver, CO & 60 & 72 & 28 (58.3) & 24 (36.9) \\
Honolulu, HI & 63 & 74 & 19 (37.3) & 27 (39.7) \\
Houston, TX & 48 & 167 & 48 (43.2) & 54 (38.6) \\
Las Vegas, NV & 40 & 179 & 43 (38.7) & 57 (38.8) \\
Los Angeles, CA & 66 & 224 & 77 (48.7) & 62 (31.3) \\
Miami, FL & 78 & 98 & 21 (39.6) & 32 (37.6) \\
New Orleans, LA & 57 & 70 & 26 (56.5) & 13 (20.6) \\
New York, NY & 89 & 257 & 90 (55.2) & 74 (32.0) \\
Orlando, FL & 41 & 195 & 57 (44.9) & 72 (42.1) \\
Philadelphia, PA & 78 & 75 & 26 (54.2) & 20 (32.3) \\
Phoenix, AZ & 40 & 97 & 32 (43.2) & 38 (42.2) \\
Portland, OR & 64 & 72 & 30 (55.6) & 19 (31.1) \\
San Antonio, TX & 36 & 102 & 25 (36.8) & 32 (35.6) \\
San Diego, CA & 50 & 196 & 57 (47.5) & 54 (32.1) \\
San Francisco, CA & 86 & 205 & 57 (42.5) & 61 (33.3) \\
San Jose, CA & 50 & 95 & 31 (51.7) & 21 (24.7) \\
Seattle, WA & 73 & 103 & 29 (50.0) & 24 (26.4) \\
\bottomrule
\end{tabular}
\label{table:table_location_walkability}
\end{table}

} %

\end{document}